\documentclass[%
 aip,
 pof,
 amsmath,amssymb,
 preprint,%
]{revtex4-1}

\usepackage[utf8]{inputenc}
\usepackage[T1]{fontenc}
\usepackage{mathptmx}
\usepackage{graphicx}
\usepackage{dcolumn}
\usepackage{bm}
\usepackage{booktabs}
\usepackage{siunitx}
\usepackage{xcolor}
\usepackage[section]{placeins}
\usepackage{hyperref}
\usepackage{cleveref}
\usepackage{orcidlink}
\hypersetup{hidelinks}
\usepackage{etoolbox}
\makeatletter
\def\@email#1#2{%
 \endgroup
 \patchcmd{\titleblock@produce}
  {\frontmatter@RRAPformat}
  {\frontmatter@RRAPformat{\produce@RRAP{*#1\href{mailto:#2}{#2}}}\frontmatter@RRAPformat}
  {}{}
}%
\makeatother

\graphicspath{{figures/}}

\newcommand{\LP}{\mathrm{LP}}

\newcommand{\Rn}{R_{ij}^{\vert n}}

\newcommand{\calRnk}{\mathcal{R}_{ij}^{\vert n,k}}
\newcommand{\calKnk}{\mathcal{K}^{\vert n,k}}

\newcommand{\Psifd}{\Psi_{fd}}
\newcommand{\CSRR}{C_{\mathrm{SRR}}}
\newcommand{\Bpsi}{B_{\Psi}}
\newcommand{\kLP}{\kappa_{\mathrm{LP}}}
\newcommand{\pospart}[1]{\left[#1\right]_+}

\begin{document}

\preprint{Accepted Manuscript}

\title[Ray--Column IPRM]{Scale-conditioned structure-based closure for homogeneous turbulence: Ray--Column Interacting Particle Representation Model}

\author{Stavros C. Kassinos\texorpdfstring{\,\orcidlink{0000-0002-3501-3851}}{}}
\email{kassinos@ucy.ac.cy}
\homepage{https://ucy-compsci.org}
\affiliation{Computational Sciences Laboratory, University of Cyprus, 1 University Avenue, 2109 Aglantzia, Nicosia, Cyprus}

\date{July 2026}

\begin{abstract}
The particle representation model (PRM) and the interacting particle representation model (IPRM) represent homogeneous turbulence through an ensemble of orientation-conditioned structural states.  In the classical cluster-averaged form, the conditional state is organized by the unit spectral direction, while the radial spectral coordinate has already been integrated out.  This paper introduces a scale-conditioned extension in which the spectral vector is decomposed into orientation and radial wavenumber, and the conditional structure state is projected onto finite radial bands.  We refer to the resulting finite-band formulation as Ray--Column IPRM, or RC-IPRM.

The formulation is developed first at the level of the continuum spectral tensor and is then reduced to the ray--packet ensemble sums used in the numerical implementation.  The finite bands are interpreted as projections of an orientation--wavenumber spectral density.  Their role is to retain scale-conditioned structural populations for closure evaluation, while cascade modeling is left outside the present reference closure.  The rapid dynamics remain ray--packet resolved.  The nonlinear slow and terminal closure coefficients are evaluated from band-aggregate structure tensors, obtained by integrating over orientation and wavenumber within each retained band before the final one-point reconstruction is formed.

A reference scale-conditioned closure is obtained by combining PRM rapid kinematics, band-local effective-gradient response and slow rotational randomization based on band-aggregate structure tensors, together with an active large-scale-enstrophy (LSE) terminal-drain map.  The original IPRM used a modified $\epsilon$ equation for the second scale; in the present active-LSE closure, the continuous complementarity invariant $\Psi_{fd}$ regularizes the active LSE structure-to-dissipation map of \textit{Structure and Scales in Turbulence Modeling}.  The Ray--Column formulation then evaluates this active map on band-aggregate structural populations.  The model is assessed in irrotational strain, homogeneous shear, elliptic-streamline, and rotating-shear configurations.  The results show that the retained bands provide more than internal
resolution: they reveal how scale-conditioned structure affects spectral
migration, terminal-drain assignment, and filtered observables before
one-point averaging removes radial scale information.
\end{abstract}

\maketitle

\begin{center}
\small
This article has been accepted by \emph{Physics of Fluids}. After publication, the version of record will be available from AIP Publishing.
\end{center}

\section{Introduction}
\label{sec:introduction}

\subsection{Motivation and scope}
\label{subsec:scope}

Traditional one-point turbulence closures deliberately compress the
state of the turbulent field.  This compression is essential for
engineering prediction, but it also removes information that may control
the response of the turbulence when the mean deformation is rapid,
rotating, or far from equilibrium.  The Reynolds-stress tensor
\(R_{ij}\) carries the componentality of the turbulent kinetic energy,
but it does not by itself specify the dimensionality or circulicity of
the underlying energy-containing structures.  Structure-based models
augment the one-point description by retaining additional tensorial
information about the geometry of the turbulence structure.  The final
description is still compact, but it carries more information than the
Reynolds stress alone.

The original Particle Representation (PRM) and Interacting Particle Representation (IPRM) models  were developed from the opposite end of this
compromise~\citep{KassinosReynolds1995TF61,KassinosReynolds1996IPRM}.
They introduced an ensemble of conditional structural states whose
average recovers the one-point turbulence structure tensors~\citep{KRR2001}, including the turbulence dimensionality $D_{ij}$ and circulicity $F_{ij}$.  Thus the
model begins from structural populations and only later performs the
one-point compression.  In this way PRM/IPRM retains directional
information that is invisible in a conventional one-point state, while
avoiding the cost and complexity of a full spectral closure.  The PRM
provides an efficient representation of the rapid-distortion evolution
of homogeneous turbulence under general mean deformations.  The IPRM
adds a structure-based closure for the slow nonlinear redistribution
dynamics. 
In homogeneous turbulence the triple-correlation transport terms vanish
from the exact one-point Reynolds-stress equation.  The familiar
one-point manifestations of the slow nonlinear interactions are then the
slow pressure--strain-rate redistribution and the terminal dissipation.
IPRM does not model these effects by adding separate one-point closures.
Instead, it postulates effective gradients acting on the structural
ensemble, with incompressibility determining the associated
pressure-like response.  Together with slow rotational randomization and
a matched terminal time scale, these terms provide a combined
structure-based representation of slow redistribution and terminal
energy drain.

In the original IPRM, the inherited slow response acts on conditional
structural states and therefore influences the turbulence before the
one-point average is formed.  The effective-gradient terms modify the
gradients experienced by the structural elements, while slow rotational
randomization redistributes orientation-conditioned structure in a
trace-preserving manner.  A matched terminal time scale, supplied through
a second turbulence-scale equation, sets the rate at which the high-Re
structural model loses energy to unresolved terminal scales.  This
combination gives the IPRM much of its robustness: the one-point
structure tensors determine the slow response, but the response is
applied to the conditional structural ensemble before the final average
is taken. With suitable modeling assumptions, the IPRM was designed to respect the
realizability constraints inherited from the underlying PRM
representation. 

For the purpose of the present work, the classical IPRM leaves three
issues unresolved.  First, in the classical cluster representation the
state is conditioned on the unit vector
\begin{equation}
  n_i=\frac{N_i}{|\bm N|}.
  \label{eq:intro-n-def}
\end{equation}
This choice is computationally efficient, but the magnitude of the same
spectral vector has already been averaged out.  The model therefore
knows about turbulence structure, but it carries limited information
about turbulence scale.

Second, because the radial spectral coordinate is not retained, the
classical IPRM remains a high-Reynolds-number closure that relies on an
externally supplied second-scale equation to provide the terminal
dissipation of turbulent kinetic energy.  In the original IPRM this role
was played by a modified \(\epsilon\) equation, including a
structure-based correction for strongly rotated turbulence.  That choice
was consistent with the modeling practice of the time, but it does not
make direct use of a retained distribution of energy across radial
spectral scale, since that distribution is absent from the classical
IPRM state.

Third, the slow response in the original model is conditioned in its
application, while the tensorial closure forms and time scales used in
that response are constructed from the \emph{global one-point} structure
tensors.  Effective-gradient products such as \(r_{ik}d_{kj}\), the
corresponding scalar contractions, and the matched terminal time scale
are formed from the scale-integrated state before being applied to the
conditional structural ensemble.  Since the radial spectral coordinate
has already been averaged out, the response cannot vary separately among
different wavenumber populations.

The present paper has a specific purpose: to restore radial spectral
scale to the PRM/IPRM conditional state without sacrificing the reduced
particle character that makes IPRM attractive as an alternative to full
spectral closures.  The inclusion of radial scale also motivates a more
direct structure-based treatment of the second turbulence scale.  In the
Ray--Column IPRM developed here, the radial coordinate is retained in
finite wavenumber bands, and the Large-Scale Enstrophy (LSE) equation of
\emph{Structure and Scales in Turbulence Modeling} \citep{ReynoldsLangerKassinos2002} is adapted into an
active terminal-drain map.  The Ray--Column formulation evaluates this
map on band-aggregate structural populations.  With the slow-strain
correction introduced below, this gives a structure--scale formulation
within the IPRM line of closures.

In the terminology of the original PRM/IPRM
formulation~\citep{KassinosReynolds1995TF61,KassinosReynolds1996IPRM,KassinosLangerHaireReynolds2000, KRR2001},
the Reynolds-stress tensor carries information about componentality, the
dimensionality tensor carries information about the spatial
dimensionality of the underlying structure, and the circulicity tensor
carries the large-scale circulation tied to that structure.  In the
classical formulation these one-point tensors are obtained after
averaging over the structural ensemble.  The Ray--Column extension
preserves the same distinction across radial scale.  The ray--packet
population remains available for rapid evolution and projection, while
the slow and terminal closure coefficients are formed from
orientation- and wavenumber-integrated band populations before those
bands are summed into the global one-point state.

We call the resulting finite-band construction Ray--Column IPRM, or
RC--IPRM.  The name is descriptive: rays denote the retained orientation
directions, while columns denote the finite radial wavenumber intervals
attached to those directions.  Here we construct a scale-conditioned
structural ensemble in which effective gradients, slow rotational
randomization, and terminal energy-drain shares can be evaluated
separately on band-aggregate populations before scale information is
lost in the one-point average.  The closure is intended as a reduced
particle representation in which turbulence scale effects can influence
the evolution of componentality, dimensionality, and circulicity through
band-local, band-aggregate structural states.  Resolved conservative
cascade dynamics are left for future extensions.

This distinction becomes especially useful when the comparison
observable is itself scale dependent.  Bardina's rotating-shear data, for
example, are filtered LES quantities~\citep{BardinaFerzigerReynolds1983}.
A purely one-point IPRM reconstruction has already lost the radial
spectral coordinate needed to form a fixed low-pass observable.  In
RC--IPRM the band energies remain available, so a prescribed low-pass
quantity can be formed from retained band contributions.  This feature
makes the band representation more than internal bookkeeping, while the
model itself remains a homogeneous structural closure.

In summary, the fluid-physics point of the Ray--Column extension is that
radial scale affects how componentality, dimensionality, and circulicity
evolve under mean deformation.  Once the radial coordinate has been
averaged out, the model can still describe the global one-point
structure, but it cannot distinguish how different scale populations
contribute to slow structural response, terminal drain, or filtered
observables.  Finite radial bands expose this structure--scale coupling
while preserving the reduced particle character of PRM/IPRM.  In this
representation the effective gradients modify not only the rate of
structural distortion, but also the spectral trajectories, band
residency, and band migration of the modeled structural populations.
The active-LSE analysis also exposes a slow-strain failure mode of the
original LSE structure-to-dissipation map, in which dimensionality and
circulicity remain anisotropic but become poorly aligned for terminal
transfer.  In rotating shear, the retained band energies provide the
additional physical information needed to form filtered or low-pass
observables before one-point averaging removes the radial spectral
coordinate.

\subsection{Notation convention}
\label{subsec:notation-convention}

We use the notation of \emph{Structure and Scales in Turbulence
Modeling}~\citep{ReynoldsLangerKassinos2002} wherever it affects the
LSE formulation.  Turbulent kinetic energy is denoted by
\begin{equation}
  \kappa = \frac{1}{2}R_{ii},
  \label{eq:tke-def}
\end{equation}
following the convention of that paper.  The symbol \(k\) is reserved
for wavenumber.  Thus, for a spectral vector \(N_i\),
\begin{equation}
  n_i=\frac{N_i}{|\bm N|},
  \qquad
  k=|\bm N|.
  \label{eq:n-k-def}
\end{equation}
The scale-conditioned state is written as \(R_{ij}^{\vert n,k}\) to denote conditioning on both $n$ and $k$. 
Its projection onto a finite radial interval \(I_\beta\) is written as 
\(R_{ij}^{\vert n,\beta}\).  Band energies are denoted by
\(\kappa_\beta\); the symbol \(k_\beta\) is reserved for a wavenumber,
or for a representative band-center wavenumber when such a quantity is
needed.

\subsection{Contributions}
\label{subsec:motivation-contributions}

The first contribution is a continuum formulation in which the Reynolds
stress is reconstructed from an orientation--wavenumber spectral density.
This formulation defines the formal object \(R_{ij}^{\vert n,k}\),
explains why the velocity spectrum tensor is symmetrized before
positive-ray integrals are formed, and identifies the resulting
orientation--wavenumber object as a tensor-valued density.

The second contribution is the finite-band projection
\(R_{ij}^{\vert n,\beta}\), together with the band-aggregate tensors
obtained by integrating over both orientation and the finite wavenumber
interval.  This projection preserves the exact PRM/RDT limit when the
rapid operator is summed over bands, while exposing the current-band
boundary crossing induced by rapid wavenumber drift.

The third contribution is the explicit connection between this projected
continuum picture and the implemented ray--packet ensemble sums.  The
implementation carries weighted packets with precomputed initial
spectral-cell moments, advances their current wavenumbers, and obtains
band moments by projection of the carried packet population.

The fourth contribution is a reference scale-conditioned IPRM closure.
It combines PRM rapid kinematics with slow and terminal ingredients
evaluated from band-aggregate structure tensors, together with an active
LSE terminal-drain map.  This last choice is historically significant:
the original IPRM calculations used a modified \(\epsilon\) equation for
the second scale, while the present closure uses the LSE formulation of
\citet{ReynoldsLangerKassinos2002} as an active terminal-drain closure.
Used actively in this way, the original LSE structure factor
\(\chi=3f:d\) becomes too fragile in some slow-strain states.  We
therefore introduce a continuous complementarity invariant, \(\Psi_{fd}\),
constructed from the deviatoric circulicity and dimensionality tensors.
The correction belongs to the active LSE map within the present
RC--IPRM closure.

The final contribution is an assessment of the resulting reference model
in a set of canonical homogeneous flows.  The main figures emphasize
irrotational strain, homogeneous shear, elliptic-streamline behavior,
and rotating homogeneous shear, including a Bardina low-pass comparison.
An appendix atlas reports tensor and scalar comparisons for the full irrotational-strain family from the same calculation set; see Appendix~\ref{app:atlas}.

\section{From PRM/IPRM to Ray--Column IPRM}
\label{sec:ray-column}

\subsection{Classical structure tensors}
\label{subsec:one-point-definitions}

We recall first the structure-based notation used in PRM and IPRM in the limit of homogeneous turbulence~\citep{KassinosReynolds1995TF61,KassinosReynolds1996IPRM,KRR2001}.   Each structural element carries a fluctuating velocity $V_i$ and an orientation, or gradient, vector $n_i$.  For an incompressible structural element,
\begin{equation}
  V_i n_i=0.
  \label{eq:incompressible-element}
\end{equation}
The Reynolds-stress tensor is
\begin{equation}
  R_{ij}=\langle V_i V_j\rangle,
  \label{eq:Rij-def}
\end{equation}
where the angled brackets denote ensemble averaging.  This tensor describes the componentality of the turbulence.  The dimensionality tensor is
\begin{equation}
  D_{ij}=\langle V^2 n_i n_j\rangle,
  \label{eq:Dij-def}
\end{equation}
and describes the spatial dimensionality of the underlying structure.  The third structure tensor is the circulicity tensor,
\begin{equation}
F_{ij}
=
\left\langle
\left(\epsilon_{ik\ell}V_k n_\ell\right)
\left(\epsilon_{jmn}V_m n_n\right)
\right\rangle .
 \label{eq:Fij-def}
\end{equation}
Equivalently, if
\begin{equation}
s_i=\frac{\epsilon_{ijk}V_j n_k}{V}.
\label{eq:streamfunction-direction}
\end{equation}
is the normalized streamfunction direction associated
with the structural element, then
\begin{equation}
F_{ij}=\langle V^2s_is_j\rangle .
 \label{eq:Fij-m-def}
\end{equation}
Thus $F_{ij}$ carries the large-scale circulation tied to the turbulence structure.

The traces of the three structure tensors are equal,
\begin{equation}
  R_{kk}=D_{kk}=F_{kk}=\langle V^2\rangle=q^2=2\kappa.
  \label{eq:equal-traces}
\end{equation}
For each incompressible structural element, the componentality, dimensionality, and circulicity directions form an orthogonal structural triad, and hence
\begin{equation}
  R_{ij}+D_{ij}+F_{ij}=q^2\delta_{ij}.
  \label{eq:RDF-identity}
\end{equation}
The corresponding normalized tensors are
\begin{equation}
  r_{ij}=\frac{R_{ij}}{q^2},\qquad
  d_{ij}=\frac{D_{ij}}{q^2},\qquad
  f_{ij}=\frac{F_{ij}}{q^2},
  \label{eq:rdf-normalized}
\end{equation}
so that
\begin{equation}
  r_{ij}+d_{ij}+f_{ij}=\delta_{ij}.
  \label{eq:rdf-normalized-identity}
\end{equation}
Their deviatoric parts are denoted by
\begin{equation}
  \widetilde r_{ij}=r_{ij}-\frac13\delta_{ij},\qquad
  \widetilde d_{ij}=d_{ij}-\frac13\delta_{ij},\qquad
  \widetilde f_{ij}=f_{ij}-\frac13\delta_{ij},
  \label{eq:rdf-deviators}
\end{equation}
and satisfy
\begin{equation}
  \widetilde r_{ij}+\widetilde d_{ij}+\widetilde f_{ij}=0.
  \label{eq:rdf-deviator-identity}
\end{equation}
These definitions are the standard structure-based componentality, dimensionality, and circulicity notation used throughout this modeling line~\citep{KassinosReynolds1995TF61,KassinosLangerHaireReynolds2000,KRR2001}.

\subsection{Original cluster PRM/IPRM}
\label{subsec:prm-one-point}

In the original cluster-averaged PRM/IPRM formulation~\citep{KassinosReynolds1996IPRM} the elementary conditional label is the direction $n_i=N_i/|\bm N|$.  The conditional stress associated with a single cluster may be written schematically as
\begin{equation}
  \Rn = \left\langle V_i V_j \mid n\right\rangle,
  \label{eq:orig-cond-stress}
\end{equation}
and the one-point Reynolds stress is recovered by averaging over the retained orientation set.  In the equal-orientation cluster implementation this reconstruction has the form
\begin{equation}
  R_{ij} \approx \frac{1}{N_n}\sum_{a=1}^{N_n} R_{ij}^{\vert n_a}.
  \label{eq:orig-cluster-average}
\end{equation}
Equation~\eqref{eq:orig-cluster-average} should be read as a simple equal-orientation cluster average.  The factor $1/N_n$ is the normalization of this finite average.  It may be viewed formally as a uniform quadrature factor, but its role here is purely the normalization of the cluster average.  The missing piece, for the purposes of the present work, is the radial coordinate associated with the same spectral vector $N_i$.

\subsection{Continuum orientation--wavenumber formulation}
\label{subsec:continuum-formulation}

The scale-conditioned extension is most cleanly defined before discretization.  Let \(\Phi_{ij}(\bm N,t)\) denote the symmetrized velocity spectrum tensor, normalized so that
\begin{equation}
  R_{ij}(t)
  =
  \int_{\mathbb R^3}
  \Phi_{ij}(\bm N,t)\,\mathrm d^3\bm N .
  \label{eq:spectral-tensor-reconstruction}
\end{equation}
The word ``symmetrized'' is used here in the structure-tensor sense.  A directed Fourier correlation restricted to a single positive ray can produce a tensor with an antisymmetric part, whereas the PRM/IPRM conditional objects are Reynolds-stress-like structure tensors.  They are therefore formed from the symmetric contribution associated with the spectral ray, equivalently by pairing the contributions from the opposed Fourier directions before the positive radial coordinate is introduced.  With this convention
\begin{equation}
  \Phi_{ij}(\bm N,t)=\Phi_{ji}(\bm N,t),
  \label{eq:Phi-symmetric}
\end{equation}
and the ray and band integrals below define symmetric conditional stress tensors.  The symmetrization is a spectral bookkeeping convention used to define Reynolds-stress-like conditional tensors.

We now decompose the spectral vector into orientation and radial wavenumber,
\begin{equation}
  \bm N=k n,
  \qquad |n|=1,
  \qquad k=|\bm N|\ge 0 .
  \label{eq:N-kn-decomp}
\end{equation}
Since \(\mathrm d^3\bm N = k^2\,\mathrm dk\,\mathrm d\Omega_n\), it is useful to absorb the Jacobian into the orientation--wavenumber tensor density
\begin{equation}
  \mathcal R_{ij}^{\vert n,k}(t)=k^2\Phi_{ij}(k n,t).
  \label{eq:calR-def}
\end{equation}
The one-point stress is then reconstructed as
\begin{equation}
  R_{ij}(t)=\int_0^\infty \int_{\mathbb S^2}
  \calRnk(t)\,d\Omega_n\,dk.
  \label{eq:calR-reconstruction}
\end{equation}
The use of the full orientation sphere in Eq.~\eqref{eq:calR-reconstruction} is a convenient continuum convention.  Some implementations represent orientations by equal clusters or by unoriented ray pairs; the corresponding factors then belong to the normalization of the finite average.

The associated energy density is
\begin{equation}
  \calKnk(t)=\frac{1}{2}\mathcal{R}_{ii}^{\vert n,k}(t),
  \qquad
  \kappa(t)=\int_0^\infty \int_{\mathbb S^2}
  \calKnk(t)\,d\Omega_n\,dk.
  \label{eq:energy-density}
\end{equation}
The tensor density \(\calRnk\) is a tensor-valued spectral density.  A probabilistic interpretation would use the non-negative energy density \(\calKnk\), normalized by \(\kappa\), as the corresponding measure.  The formulation below uses the tensor density directly.

\subsection{Finite band projection}
\label{subsec:band-projection}

The Ray--Column state used in the computations is a finite projection of the continuum object just defined.  For a radial wavenumber interval $I_\beta$, define
\begin{equation}
  R_{ij}^{\vert n,\beta}(t)
  =\int_{I_\beta}\mathcal{R}_{ij}^{\vert n,k}(t)\,dk.
  \label{eq:band-projection}
\end{equation}
Equation~\eqref{eq:band-projection} is the continuum definition of the band projection.  In the computations reported below the finite radial packet quadrature of Sec.~\ref{subsec:implementation-sums} realizes this projection: each packet carries an initial spectral-cell weight and is assigned to a current band according to its evolved wavenumber.
The one-point tensor follows from the sum over bands and the integral over orientations,
\begin{equation}
  R_{ij}(t)=\sum_{\beta}\int_{\mathbb S^2}
  R_{ij}^{\vert n,\beta}(t)\,d\Omega_n.
  \label{eq:band-reconstruction}
\end{equation}
The band energies are
\begin{equation}
  \kappa_\beta(t)=\int_{\mathbb S^2}\frac{1}{2}R_{ii}^{\vert n,\beta}(t)\,d\Omega_n,
  \qquad
  \kappa(t)=\sum_\beta \kappa_\beta(t).
  \label{eq:band-energy}
\end{equation}
The bands are therefore projections of an underlying orientation--wavenumber description.  Their boundaries are numerical projection choices, and their purpose is to retain enough radial information for structural operations and filtered observables to be defined before the one-point average is taken.  Universal inertial-range shells or physical band constants would belong to a different model specification.

\subsection{RDT consistency and current-band crossing}
\label{subsec:rdt-consistency-current-crossing}

The finite-band projection leaves the rapid PRM/RDT kinematics intact.  Here and below
\(G_{ij}=\partial U_i/\partial x_j\) denotes the imposed mean-velocity gradient and
\(S_{ij}=(G_{ij}+G_{ji})/2\) its symmetric part.  In homogeneous rapid distortion the spectral vector evolves as a covector.  With
\begin{equation}
  \xi=\log k,
\end{equation}
the radial drift is
\begin{equation}
  \dot\xi=\frac{\dot k}{k}
  =-G_{ij}n_i n_j
  =-S_{ij}n_i n_j,
  \label{eq:xi-dot-main}
\end{equation}
because the antisymmetric part of the mean gradient drops out of the double contraction with $n_i n_j$.  Thus a band defined by current wavenumber can exchange content with neighboring bands even in inviscid RDT.  This exchange is a rapid kinematic boundary-crossing effect caused by the wavevector evolution.

The implementation used in the present calculations handles this rapid boundary flux by projection.  The ray--packet quadrature carries the evolving orientation, velocity covariance, and radial shift, and band quantities are obtained by projecting the current packet population onto the chosen wavenumber intervals.  In an equivalent finite-volume description of the same projection, boundary terms appear at the band edges and telescope when all bands are summed.  Therefore the summed Ray--Column RDT evolution recovers the original ray-integrated PRM/RDT equation.  The departures introduced by the reference RC-IPRM closure enter through the band-aggregate slow response and terminal-drain terms while the rapid/RDT operator is retained.  The short derivation is given in Appendix~\ref{app:rdt-consistency}.

\subsection{Ray--band and band-aggregate structure tensors}
\label{subsec:rc-iprm-conditional-definitions}

The same structure-tensor definitions apply to projected populations,
but two projection levels must be kept separate.  The first level is the
orientation-resolved band projection.  For a finite radial interval
\(I_\beta\), the stress projection has already been defined in
Eq.~\eqref{eq:band-projection}; the corresponding dimensionality and
circulicity projections are
\begin{equation}
  D_{ij}^{\vert n,\beta}
  =\int_{I_\beta}\mathcal D_{ij}^{\vert n,k}\,dk,
  \qquad
  F_{ij}^{\vert n,\beta}
  =\int_{I_\beta}\mathcal F_{ij}^{\vert n,k}\,dk .
  \label{eq:ray-band-DF}
\end{equation}
Here \(\mathcal D_{ij}^{\vert n,k}\) and
\(\mathcal F_{ij}^{\vert n,k}\) are the orientation--wavenumber tensor
densities corresponding to dimensionality and circulicity. The ray--band tensors \(R_{ij}^{\vert n,\beta}\),
\(D_{ij}^{\vert n,\beta}\), and \(F_{ij}^{\vert n,\beta}\) describe how
componentality, dimensionality, and circulicity are distributed across
orientation and projected radial scale, where the latter is represented
by finite wavenumber bands.  They are useful formal and diagnostic
objects.

The closure state used in the reference model is one level more
aggregated.  For each band, define the orientation- and
wavenumber-integrated band tensors
\begin{equation}
  R_{ij}^{\beta}
  =\int_{\mathbb S^2}\int_{I_\beta}
  \mathcal R_{ij}^{\vert n,k}\,dk\,d\Omega_n,
  \qquad
  D_{ij}^{\beta}
  =\int_{\mathbb S^2}\int_{I_\beta}
  \mathcal D_{ij}^{\vert n,k}\,dk\,d\Omega_n,
  \qquad
  F_{ij}^{\beta}
  =\int_{\mathbb S^2}\int_{I_\beta}
  \mathcal F_{ij}^{\vert n,k}\,dk\,d\Omega_n .
  \label{eq:band-integrated-rdf}
\end{equation}
These tensors are local to the band index \(\beta\), but they have
already been averaged over both the orientation sphere and the
wavenumber interval of that band.  The one-point structure tensors are
recovered by summing the band tensors,
\begin{equation}
  R_{ij}=\sum_\beta R_{ij}^{\beta},\qquad
  D_{ij}=\sum_\beta D_{ij}^{\beta},\qquad
  F_{ij}=\sum_\beta F_{ij}^{\beta}.
  \label{eq:band-sum-rdf}
\end{equation}

For each nonempty band, the common trace is
\begin{equation}
  q_\beta^2
  =R_{kk}^{\beta}
  =D_{kk}^{\beta}
  =F_{kk}^{\beta},
  \label{eq:qbeta-trace}
\end{equation}
and the normalized band tensors are
\begin{equation}
  r_{ij}^{\beta}=\frac{R_{ij}^{\beta}}{q_\beta^2},\qquad
  d_{ij}^{\beta}=\frac{D_{ij}^{\beta}}{q_\beta^2},\qquad
  f_{ij}^{\beta}=\frac{F_{ij}^{\beta}}{q_\beta^2}.
  \label{eq:band-rdf-normalized}
\end{equation}
The global normalized tensors are energy-weighted averages of these
band-normalized tensors,
\begin{equation}
  r_{ij}=\sum_\beta \theta_\beta r_{ij}^{\beta},\qquad
  d_{ij}=\sum_\beta \theta_\beta d_{ij}^{\beta},\qquad
  f_{ij}=\sum_\beta \theta_\beta f_{ij}^{\beta},
  \qquad
  \theta_\beta=\frac{q_\beta^2}{q^2}.
  \label{eq:global-from-band-local}
\end{equation}
Thus the band reconstruction identity is algebraic and exact for the
projected state.  The modeling choice enters later, when nonlinear slow
and terminal closure operations are evaluated from the band-aggregate
normalized tensors in Eq.~\eqref{eq:band-rdf-normalized}.

Formal normalized ray--band tensors may also be introduced whenever the
ray--band trace is nonzero, for example
\(r_{ij}^{\vert n,\beta}=R_{ij}^{\vert n,\beta}/q^{2\vert n,\beta}\).
These quantities diagnose orientation-resolved structure inside a band.
They are not the nonlinear closure variables used in the reference
calculations.

\subsection{Implementation as ray--packet ensemble sums}
\label{subsec:implementation-sums}

The preceding equations define continuum and band-projected objects.  In
the computations these projections are realized by a finite ray--packet
quadrature.  The orientation sphere is represented by \(N_r\) nearly
equal-area ray directions \(n_a\), \(a=1,\ldots,N_r\), with orientation
quadrature normalization
\begin{equation}
  \Omega_a=\frac{1}{N_r}.
  \label{eq:orientation-normalization}
\end{equation}
Each ray carries the same radial packet grid.  The packet index
\(q=1,\ldots,N_k\) labels an interval \(J_q\) in the initial wavenumber
coordinate \(k_0\).  The representative initial wavenumber of this
interval is denoted by \(k_{0q}\).  The prescribed initial scalar energy
spectrum is \(E_0(k_0)\), normalized to the chosen initial turbulent
kinetic energy,
\[
  \int_0^\infty E_0(k_0)\,dk_0=\kappa_0 .
\]
For each radial packet we precompute the initial spectral-cell moments
\begin{equation}
  \mathcal I_{mq}^{(0)}
  =
  \int_{J_q} k_0^m E_0(k_0)\,dk_0,
  \qquad m=0,2 .
  \label{eq:packet-spectral-weights}
\end{equation}
Thus \(\mathcal I_{0q}^{(0)}\) is the packet energy quadrature measure,
while \(\mathcal I_{2q}^{(0)}\) is the corresponding
\(k_0^2\)-weighted measure used for large-scale-enstrophy moments.  The
superscript \((0)\) indicates that these are fixed measures of the
initial spectrum.  They do not evolve; the packet evolution is carried
by the variables described below.

The carried variables for ray--packet \((a,q)\) include the packet
orientation \(n_i^{(aq)}(t)\), the velocity covariance tensor
\(C_{ij}^{(aq)}(t)\), and a logarithmic radial shift \(s^{(aq)}(t)\).
The shift is defined by
\begin{equation}
  s^{(aq)}(t)
  =
  \log\frac{k^{(aq)}(t)}{k_{0q}},
  \qquad
  k^{(aq)}(t)=k_{0q}\exp s^{(aq)}(t),
  \label{eq:packet-current-k}
\end{equation}
so that \(s^{(aq)}=0\) initially.  Thus \(s^{(aq)}\) records the current
radial displacement of the packet in wavenumber space.  The tensor
\(C_{ij}^{(aq)}\) carries the packet velocity covariance or
amplification; its trace \(C_{\ell\ell}^{(aq)}\) gives the packet
velocity-variance factor entering the dimensionality tensor.

For a current band \(I_\beta=[k_\beta^-,k_\beta^+)\), define the
membership indicator
\begin{equation}
  \mathbf{1}_{\beta}^{(aq)}(t)=
  \begin{cases}
    1, & k^{(aq)}(t)\in I_\beta,\\
    0, & \text{otherwise}.
  \end{cases}
  \label{eq:packet-band-indicator}
\end{equation}
A finite band therefore contains the subset of ray--packets whose
current wavenumbers lie in that interval,
\[
  \mathcal P_\beta(t)
  =
  \left\{
  (a,q): k^{(aq)}(t)\in I_\beta
  \right\}.
\]
All rays are initialized with the same radial packet grid, but the
current packet population of a band need not be the same for all
\(\beta\), nor remain fixed in time.  Packets move between current
bands as their radial shifts evolve.  The packet count
\(|\mathcal P_\beta(t)|\) is therefore only a diagnostic of sampling.
The band tensors are defined by the quadrature sums below, with packet
measures \(\Omega_a\mathcal I_{0q}^{(0)}\) or the corresponding
higher-moment measures, rather than by an equal-count average over the
packets in the band.

The formal \(k\)-integrals over \(I_\beta\) are represented at runtime by
summing the quadrature contributions of the packets whose current
wavenumbers lie in that band.  The current band assignment is therefore
handled by the evolved packet wavenumber \(k^{(aq)}(t)\), while the
spectral-cell measures \(\mathcal I_{0q}^{(0)}\) and
\(\mathcal I_{2q}^{(0)}\) remain attached to the packet.

The implemented band-aggregate tensors are computed schematically as
\begin{equation}
  \widehat R_{ij}^{\beta}
  =
  \sum_{a=1}^{N_r}\sum_{q=1}^{N_k}
  \Omega_a \mathcal I_{0q}^{(0)}\,
  \mathbf{1}_{\beta}^{(aq)}
  C_{ij}^{(aq)},
  \label{eq:implemented-Rbeta}
\end{equation}
\begin{equation}
  \widehat D_{ij}^{\beta}
  =
  \sum_{a=1}^{N_r}\sum_{q=1}^{N_k}
  \Omega_a \mathcal I_{0q}^{(0)}\,
  \mathbf{1}_{\beta}^{(aq)}
  C_{\ell\ell}^{(aq)} n_i^{(aq)}n_j^{(aq)},
  \label{eq:implemented-Dbeta}
\end{equation}
and
\begin{equation}
  \widehat q_\beta^2=\widehat R_{kk}^{\beta},
  \qquad
  \widehat F_{ij}^{\beta}
  =
  \widehat q_\beta^2\delta_{ij}
  -\widehat R_{ij}^{\beta}
  -\widehat D_{ij}^{\beta}.
  \label{eq:implemented-Fbeta}
\end{equation}
Hats denote finite quadrature approximations to the continuum
band-aggregate tensors.  The construction of \(\widehat F_{ij}^{\beta}\)
uses the structure identity \(R_{ij}+D_{ij}+F_{ij}=q^2\delta_{ij}\),
applied to the band population.

The corresponding large-scale-enstrophy moment used for LSE
initialization is obtained with the second initial spectral-cell moment:
\begin{equation}
  \widehat H_{ij}^{\beta}
  =
  \sum_{a=1}^{N_r}\sum_{q=1}^{N_k}
  \Omega_a \mathcal I_{2q}^{(0)} e^{2s^{(aq)}}\,
  \mathbf{1}_{\beta}^{(aq)}
  C_{ij}^{(aq)},
  \qquad
  Z_\beta=\frac12\widehat H_{kk}^{\beta}.
  \label{eq:implemented-Hbeta}
\end{equation}
The difference between
Eqs.~\eqref{eq:implemented-Rbeta}--\eqref{eq:implemented-Dbeta}
and Eq.~\eqref{eq:implemented-Hbeta} is the order of the
current-wavenumber moment being formed.  The stress and dimensionality
tensors are zeroth moments in current wavenumber, so their packet
measure is \(\mathcal I_{0q}^{(0)}\).  The large-scale-enstrophy tensor
is a second current-wavenumber moment.  Since
\[
  \left(k^{(aq)}(t)\right)^2
  =
  k_{0q}^2 e^{2s^{(aq)}(t)},
\]
the fixed initial measure \(\mathcal I_{2q}^{(0)}\) must be multiplied
by \(e^{2s^{(aq)}}\) to represent the corresponding current
\(k^2\)-weighted moment.  The radial shift affects all band quantities
through the current membership indicator \(\mathbf 1_\beta^{(aq)}\);
only moments carrying explicit powers of current wavenumber acquire the
additional factor \(e^{m s^{(aq)}}\). The scalar \(Z_\beta\) is the band large-scale-enstrophy measure used in
the initialization of the active LSE variables.

The orientation-resolved object \(R_{ij}^{\vert n,\beta}\) is useful in
the formal development, but the reference closure coefficients in the
implementation are formed from the band-aggregate packet sums in
Eqs.~\eqref{eq:implemented-Rbeta}--\eqref{eq:implemented-Fbeta}.  The
orientation normalization \(\Omega_a\), the fixed initial spectral-cell
measures \(\mathcal I_{0q}^{(0)}\) and \(\mathcal I_{2q}^{(0)}\), and
the current band membership \(\mathbf 1_\beta^{(aq)}\) define the
numerical realization of the continuum projection.  They are quadrature
measures, not adjustable model coefficients.

For a diagnostic or closure quantity \(A^{(aq)}\), a band average over
the packet population is therefore a quadrature-normalized average.  For
an energy-weighted packet quantity, for example,
\begin{equation}
  \langle A\rangle^\beta
  =
  \frac{
  \sum_{a=1}^{N_r}\sum_{q=1}^{N_k}
  \Omega_a \mathcal I_{0q}^{(0)}
  \mathbf 1_\beta^{(aq)} A^{(aq)}
  }{
  \sum_{a=1}^{N_r}\sum_{q=1}^{N_k}
  \Omega_a \mathcal I_{0q}^{(0)}
  \mathbf 1_\beta^{(aq)}
  }.
  \label{eq:quadrature-band-average}
\end{equation}
Other diagnostics use the corresponding quadrature measure appropriate
to the moment being formed.  In an ideal equal-packet representation this
expression reduces to a simple population average.  In the present
implementation the radial spectral quadrature is carried explicitly
through the packet measures.

\paragraph{Initial ray--packet ensemble.}
The calculations are initialized from an isotropic ray--packet ensemble.
The orientation set \(\{n_a\}_{a=1}^{N_r}\) is used as an
equal-orientation quadrature, and each orientation carries the same
initial radial packet grid.  The initial scalar spectrum \(E_0(k_0)\) is
normalized so that
\begin{equation}
  \int_0^\infty E_0(k_0)\,dk_0=\kappa_0 .
  \label{eq:initial-spectrum-normalization}
\end{equation}
In the calculations reported here \(\kappa_0=1/2\), so that
\(q_0^2=2\kappa_0=1\).  The packet measures
\(\mathcal I_{mq}^{(0)}\) defined in Eq.~\eqref{eq:packet-spectral-weights}
are therefore fixed measures of the initial spectrum; the subsequent
evolution is carried by \(n_i^{(aq)}(t)\), \(C_{ij}^{(aq)}(t)\),
\(s^{(aq)}(t)\), and the current band membership
\(\mathbf 1_\beta^{(aq)}(t)\).

The initial packet covariance is chosen as the transverse projector,
\begin{equation}
  C_{ij}^{(aq)}(0)=\delta_{ij}-n_i^{(a)}n_j^{(a)},
  \label{eq:initial-packet-covariance}
\end{equation}
and the initial radial shift is zero,
\begin{equation}
  s^{(aq)}(0)=0,
  \qquad
  k^{(aq)}(0)=k_{0q}.
  \label{eq:initial-radial-shift}
\end{equation}
With the equal-orientation quadrature, this gives an isotropic initial
one-point structure,
\begin{equation}
  R_{ij}(0)=D_{ij}(0)=F_{ij}(0)
  =
  \frac{q_0^2}{3}\delta_{ij}.
  \label{eq:initial-isotropic-rdf}
\end{equation}
Thus the radial packet grid supplies the initial distribution of energy
and large-scale-enstrophy moments across \(k_0\), while the initial
componentality, dimensionality, and circulicity are isotropic before the
imposed homogeneous deformation begins.

Additional coefficients enter only when a different observable is
deliberately constructed from the retained bands.  For the Bardina
filtered comparison, prescribed low-pass coefficients \(a_\beta\) define
\begin{equation}
  \kappa_{\LP}=\sum_\beta a_\beta\kappa_\beta,
  \qquad
  \eta_\beta^{\LP}
  =
  \frac{a_\beta\kappa_\beta}
       {\sum_\gamma a_\gamma\kappa_\gamma}.
  \label{eq:lowpass-coefficients}
\end{equation}
The coefficients \(a_\beta\) define the filtered observable and are kept
separate from the unfiltered RC--IPRM reconstruction.

\subsection{Band-aggregate closure level and legacy limit}
\label{subsec:reconstruction-closure-level}

The Ray--Column projection and the closure level are distinct parts of
the model.  The projection defines ray--wavenumber and band-aggregate
structural populations.  The reference closure then uses the
band-aggregate tensors of Eq.~\eqref{eq:band-rdf-normalized} to define
the nonlinear slow and terminal ingredients.  Thus the closure is
band-local because it is evaluated separately for each \(\beta\), and
band-aggregate because its tensor products and scalar contractions are
formed after averaging over both orientation and wavenumber within
\(I_\beta\).

This ordering avoids a higher-moment closure hierarchy.  Products such
as
\begin{equation}
  r_{ik}^{\beta}d_{kj}^{\beta},\qquad
  r_{ik}^{\beta}d_{km}^{\beta}r_{mi}^{\beta},\qquad
  f_{ij}^{\beta}d_{ji}^{\beta}
  \label{eq:band-product-examples}
\end{equation}
are products of band-aggregate tensors.  A closure constructed instead
from fully ray-conditioned products such as
\(r_{ik}^{\vert n,\beta}d_{kj}^{\vert n,\beta}\), or from fully
\((n,k)\)-local products, would generate additional angular and radial
correlations when averaged back to the retained band tensors.  Those
correlations are not carried in the present reference model.  Each band
therefore behaves as a scale-conditioned structural population,
analogous to the one-point population in the original IPRM but restricted
to a finite radial wavenumber interval.

With this convention, a schematic ray-level equation for packets whose
current wavenumber lies in \(I_\beta\) may be written as
\begin{equation}
  \frac{\partial}{\partial t}\mathcal R_{ij}^{\vert n,k}
  =
  \mathcal L_{ij}\!\left[
    G;\mathcal R^{\vert n,k},n
  \right]
  +
  \mathcal C_{ij}\!\left[
    \mathcal R^{\vert n,k},n;\mathcal B^\beta
  \right],
  \label{eq:ray-equation-band-closure}
\end{equation}
where \(\mathcal L_{ij}\) is the rapid PRM/RDT operator and
\(\mathcal B^\beta\) denotes the band-aggregate closure state,
\begin{equation}
  \mathcal B^\beta
  =
  \left\{
  R_{ij}^{\beta},D_{ij}^{\beta},F_{ij}^{\beta},
  r_{ij}^{\beta},d_{ij}^{\beta},f_{ij}^{\beta},
  \omega_{L,\beta},\epsilon_\beta^*
  \right\}.
  \label{eq:band-closure-state}
\end{equation}
The notation emphasizes the split in the reference implementation: the
rapid part acts on ray--packet variables, while the slow contribution
may act on packet variables using coefficients determined from the
band-aggregate structural state.

The corresponding band equation follows by summing or integrating the
ray-level equation over the orientation sphere and the wavenumber
interval of the band,
\begin{equation}
  \frac{dR_{ij}^{\beta}}{dt}
  =
  \int_{\mathbb S^2}\int_{I_\beta}
  \mathcal L_{ij}\!\left[
    G;\mathcal R^{\vert n,k},n
  \right]
  dk\,d\Omega_n
  +
  \int_{\mathbb S^2}\int_{I_\beta}
  \mathcal C_{ij}\!\left[
    \mathcal R^{\vert n,k},n;\mathcal B^\beta
  \right]
  dk\,d\Omega_n .
  \label{eq:band-equation-band-closure}
\end{equation}
In the implemented ray--packet ensemble, the same operation is the
weighted packet sum described in Sec.~\ref{subsec:implementation-sums}.
The closure coefficients are constant with respect to \(n\) and \(k\)
inside a band, but the packet variables on which they act continue to
carry orientation, covariance, and current wavenumber.

\begin{figure}[!htbp]
  \centering
  \includegraphics[width=0.90\textwidth]{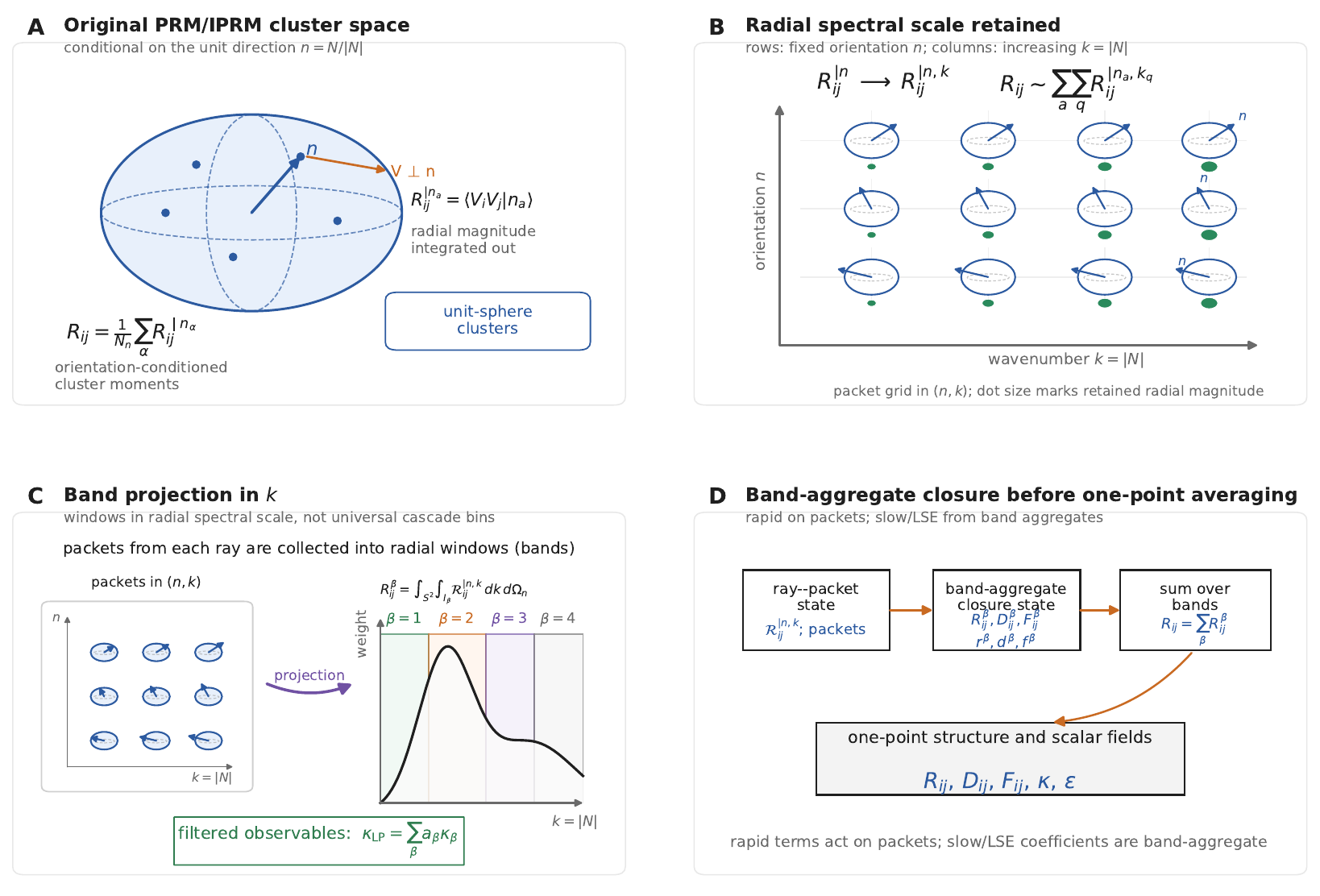}
  \caption{Continuum, band-projected, and implemented views of the Ray--Column representation.  Panel A recalls the original orientation-conditioned PRM/IPRM cluster average, in which conditional states are labeled by the direction $n=\bm N/|\bm N|$ and one-point tensors are recovered by the normalized equal-orientation average.  Panel B restores the radial coordinate $k=|\bm N|$ and shows the schematic summation over orientation rays and radial packets before band projection.  Panel C shows the projection onto finite wavenumber bands, $R_{ij}^{\vert n,\beta}=\int_{I_\beta}\mathcal R_{ij}^{\vert n,k}\,dk$, and the construction of band energies $\kappa_\beta$.  Such retained band information permits fixed low-pass observables, for example $\kappa_{\rm LP}=\sum_\beta a_\beta\kappa_\beta$, to be formed before the one-point reconstruction; the coefficients $a_\beta$ belong to the filtered observable and are kept separate from the baseline reconstruction.  Panel D summarizes the modeling order used in the reference RC--IPRM closure: the rapid PRM/RDT operator acts on ray--packet states, while effective-gradient response, slow rotational randomization, and LSE terminal-drain shares are evaluated from band-aggregate populations; the resulting band contributions are then summed to recover one-point quantities.}
  \label{fig:ray-column-schematic}
\end{figure}

The closure level also determines the relation to the legacy IPRM
limit.  If the inherited effective-gradient and slow-rotational-
randomization terms are evaluated from the global IPRM structure tensors,
the Ray--Column decomposition is a scale projection of the original model
and the summed equations recover the parent IPRM one-point evolution.
The active LSE branch is the new terminal-drain closure in the present
work; replacing it by the modified \(\epsilon\) equation used in the
original IPRM formulation~\citep{KassinosReynolds1996IPRM} recovers the
parent IPRM closure.

The reference RC--IPRM closure departs from this legacy limit by forming
the effective-gradient tensors, slow-randomization rates, and LSE
terminal-drain shares from the band-aggregate normalized tensors
\(r_{ij}^{\beta}\), \(d_{ij}^{\beta}\), and \(f_{ij}^{\beta}\).  The
rapid PRM/RDT operator remains linear in the unnormalized structure
tensors and is consistent with band summation, as discussed in
Sec.~\ref{subsec:rdt-consistency-current-crossing}.  The retained band
structure lets the slow structural response and terminal drain depend on
scale-conditioned componentality, dimensionality, and circulicity before
the final one-point average is formed.

\section{Reference scale-conditioned IPRM closure}
\label{sec:reference-closure}

\subsection{Reference RC--IPRM closure}
\label{subsec:one-reference-closure}

The Ray--Column projection and the closure model should be kept conceptually separate.  The projection introduces the scale-conditioned structural state; the closure specifies how this projected state evolves.  All calculations in this paper use one reference RC--IPRM closure.  This closure retains the rapid PRM/RDT kinematics of the original structure-based model and introduces scale conditioning only through the slow structural response and the high-Reynolds-number terminal-drain assignment.

The reference closure is defined by four choices.  First, the rapid PRM/RDT operator is retained in its original linear form and applied to the ray--packet variables.  Second, the inherited IPRM effective-gradient response is evaluated from band-aggregate structure tensors, so that the effective gradients depend on the componentality, dimensionality, and circulicity of each retained radial band.  Third, slow rotational randomization uses the same band-aggregate structural coefficients while acting on packet covariances within that band.  Fourth, the LSE equation of \citet{ReynoldsLangerKassinos2002} is used as an active structure-to-dissipation map; its band-aggregate evaluations determine the shares of the terminal drain assigned to the retained radial bands.  These operations are performed before the global one-point reconstruction, consistent with Eq.~\eqref{eq:band-equation-band-closure}.

The four parts of the closure play different physical roles.  The rapid operator gives the exact RDT limit of the structure-based representation.  The effective gradients modify the gradients experienced by the structural packets and therefore alter the rate of distortion, the orientation dynamics, and the current-band residency of the packet population.  The slow rotational-randomization term redistributes packet structure in a trace-preserving manner.  The LSE branch supplies the high-Reynolds-number terminal loss associated with the unresolved dissipative end of the model.  Thus the active LSE map assigns the terminal-drain shares, while the matched IPRM time scales realize those shares through band-aggregate effective-gradient and slow-randomization coefficients.

In the calculations reported here, the band LSE variables are initialized so that the initial raw LSE drain shares agree with the corresponding large-scale-enstrophy shares.  This initialization fixes the initial distribution of the active LSE terminal-drain map while preserving the reconstruction identities of the Ray--Column projection and the exact rapid/RDT limit.

\subsection{Original high-Reynolds-number LSE form}
\label{subsec:original-lse-form}

A careful historical distinction is needed.  The original IPRM calculations used a modified $\epsilon$ equation as the second-scale equation~\citep{KassinosReynolds1996IPRM}.  The LSE equation was introduced by \citet{ReynoldsLangerKassinos2002} as a structure-sensitive equation for the large-scale enstrophy of the energy-containing motions.  The present paper activates that LSE relation as the terminal-drain branch of a Ray--Column closure.

In the homogeneous high-Reynolds-number form relevant here, and after absorbing the energy-transfer constant into the definition of the large-scale vorticity scale $\omega_L$, the LSE model may be written
\begin{align}
  \frac{d\kappa}{dt}
  &=-2\kappa r_{ij}S_{ji}-\epsilon_{\rm LSE},
  \label{eq:lse-original-kappa}\\[0.35em]
  \epsilon_{\rm LSE}
  &=\chi\,\kappa\,\omega_L,
  \qquad
  \chi=3f_{ij}d_{ji},
  \label{eq:lse-original-epsilon}\\[0.35em]
  \frac{d\omega_L}{dt}
  &=f_{ij}S_{ij}\,\omega_L
    -\left(C_T^*-C_P^*\phi\right)\omega_L^2,
  \label{eq:lse-original-omega}\\[0.35em]
  \phi&=9r_{ij}d_{jk}f_{ki}.
  \label{eq:lse-original-phi}
\end{align}
Here $S_{ij}=(G_{ij}+G_{ji})/2$.  The factor $\chi$ determines the large-scale energy-transfer rate, while $\phi$ modulates the large-scale enstrophy transfer coefficient.  Two limiting values of $\chi$ are especially important.  For an isotropic structural state, $r_{ij}=d_{ij}=f_{ij}=\delta_{ij}/3$, and therefore $\chi=1$.  In the ideal two-dimensional/two-component limit of the structure-based model, dimensionality and circulicity occupy complementary subspaces and $f_{ij}d_{ji}=0$, giving $\chi=0$.  The latter limit represents a shutdown of the LSE terminal-transfer map, with inter-band cascade modeling remaining outside this limiting argument.  The reference calculations use the $k^4$ low-wavenumber spectrum constants,
\begin{equation}
  C_T^*=\frac{3}{2},
  \qquad
  C_P^*=\frac{4}{5}.
  \label{eq:lse-k4-constants}
\end{equation}
The implementation uses the non-negative forms
\begin{equation}
  \chi=\pospart{3f_{ij}d_{ji}},
  \qquad
  \phi=\pospart{9r_{ij}d_{jk}f_{ki}},
  \qquad
  C_\omega=\pospart{C_T^*-C_P^*\phi},
  \label{eq:lse-code-positive-parts}
\end{equation}
so that Eq.~\eqref{eq:lse-original-omega} is evaluated as
\begin{equation}
  \dot\omega_L=(f_{ij}S_{ij})\omega_L-C_\omega\omega_L^2.
  \label{eq:lse-omega-code-form}
\end{equation}
The present homogeneous calculations use the high-Re terminal-drain branch of the LSE model, omitting the low-Reynolds-number viscous terms included in the full formulation of \citet{ReynoldsLangerKassinos2002}.

\subsection{Complementarity correction to the active LSE map}
\label{subsec:psifd-corrected-map}

When Eq.~\eqref{eq:lse-original-epsilon} is used actively over the homogeneous-flow set, the original factor $\chi=3f:d$ is too fragile in some slow-strain states.  The large-scale enstrophy scale can continue to grow while $\chi$ collapses, causing the terminal drain to become artificially weak.  This behavior belongs to active use of the LSE structure-to-dissipation map itself; legacy IPRM used a different second-scale equation.

The correction is constructed directly from the deviatoric
circulicity and dimensionality tensors,
\begin{equation}
  \widetilde f_{ij}=f_{ij}-\frac13\delta_{ij},
  \qquad
  \widetilde d_{ij}=d_{ij}-\frac13\delta_{ij}.
  \label{eq:fd-deviatoric-corrected}
\end{equation}
The complementarity factor is defined as the continuous invariant
\begin{equation}
  \Psi_{fd}
  =
  \frac34
  \left[
  \left(
  \widetilde f_{mn}\widetilde f_{mn}
  \,
  \widetilde d_{pq}\widetilde d_{pq}
  \right)^{1/2}
  -
  \widetilde f_{ij}\widetilde d_{ij}
  \right].
  \label{eq:psifd-corrected}
\end{equation}
By the Cauchy--Schwarz inequality, \(\Psi_{fd}\ge0\).  The factor
vanishes when either tensor is isotropic and is small when the two
deviatoric structures are aligned.  It becomes active when
dimensionality and circulicity are both anisotropic and complementary.
This form avoids any separate limiting convention near isotropy and
preserves the isotropic and 2D--2C limits of the LSE map.

For the reference closure the corrected LSE factor is
\begin{equation}
  \chi_{\rm eff}
  =\frac{(1+B_\Psi\Psi_{fd})\chi}
        {1+B_\Psi\Psi_{fd}\chi},
  \label{eq:chieff-corrected}
\end{equation}
so that
\begin{equation}
  \epsilon=\chi_{\rm eff}\,\kappa\,\omega_L.
  \label{eq:lse-corrected-global-epsilon}
\end{equation}
The unmodified LSE map is recovered when $B_\Psi\Psi_{fd}=0$.  The construction preserves the two anchoring limits just noted.  In the isotropic state the deviatoric parts of $d_{ij}$ and $f_{ij}$ vanish, so $\Psi_{fd}=0$ and $\chi_{\rm eff}=\chi=1$.  In the ideal two-dimensional/two-component shutdown limit, $\chi=0$ and Eq.~\eqref{eq:chieff-corrected} gives $\chi_{\rm eff}=0$ for any value of $\Psi_{fd}$.  Thus the correction preserves the original LSE limits while regularizing the intermediate slow-strain states in which both $d_{ij}$ and $f_{ij}$ are anisotropic and complementary but the raw contraction $3f:d$ becomes too small.  The parameter $B_\Psi$ controls the strength and sharpness with which the complementarity correction modifies the active LSE map.  It is a closure activation parameter rather than a constant derived from first principles.  The strain-family sensitivity shown in Fig.~\ref{fig:lse-correction-strain} indicates a robust useful range $B_\Psi\simeq24$--$30$; the representative value $B_\Psi=30$ is then held fixed for all shear, elliptic, rotating-shear, and Atlas calculations.

\subsection{Band-local LSE definitions and terminal-drain shares}
\label{subsec:lse-band-shares}

The global active LSE equation supplies the target terminal drain $\epsilon$ in Eq.~\eqref{eq:lse-corrected-global-epsilon}.  The retained bands determine how this terminal drain is distributed.  For each active band $\beta$ we form the invariants from the band-aggregate tensors
\begin{align}
  \chi_\beta
  &=\pospart{3f_{ij}^{\beta}d_{ji}^{\beta}},
  \label{eq:chi-beta}\\[0.35em]
  \phi_\beta
  &=\pospart{9r_{ij}^{\beta}d_{jk}^{\beta}f_{ki}^{\beta}},
  \label{eq:phi-beta}\\[0.35em]
  C_{\omega,\beta}
  &=\pospart{C_T^*-C_P^*\phi_\beta}.
  \label{eq:Comega-beta}
\end{align}
The band LSE scale variable evolves according to
\begin{equation}
  \dot\omega_{L,\beta}
  =\left(f_{ij}^{\beta}S_{ij}\right)\omega_{L,\beta}
  -C_{\omega,\beta}\omega_{L,\beta}^2.
  \label{eq:omega-beta-evolution}
\end{equation}
The same complementarity map is evaluated on the band-aggregate tensors, giving $\Psi_{fd,\beta}$ and
\begin{equation}
  \chi_{{\rm eff},\beta}
  =\frac{(1+B_\Psi\Psi_{fd,\beta})\chi_\beta}
        {1+B_\Psi\Psi_{fd,\beta}\chi_\beta}.
  \label{eq:chieff-beta}
\end{equation}
The raw band LSE drain is then
\begin{equation}
  \widetilde\epsilon_\beta
  =\chi_{{\rm eff},\beta}\,\kappa_\beta\,\omega_{L,\beta}.
  \label{eq:raw-band-lse-epsilon}
\end{equation}
The same limiting interpretation is applied band by band.  A band whose aggregate structure is isotropic has $\chi_\beta=1$ with vanishing complementarity factor, while a band-aggregate 2D--2C shutdown has $\chi_\beta=0$ and therefore zero raw terminal drain from that band.
These raw drains define terminal-drain shares,
\begin{equation}
  s_\beta^\epsilon
  =\frac{\widetilde\epsilon_\beta}
        {\sum_\gamma \widetilde\epsilon_\gamma},
  \qquad
  \sum_\beta s_\beta^\epsilon=1,
  \label{eq:lse-band-share-normalized}
\end{equation}
with the sum taken over active bands.  In depleted numerical bands the code applies the usual positivity and fallback safeguards, leaving the physical terminal-drain model unchanged.  

At the scalar accounting level, the terminal-drain target assigned to
band \(\beta\) is
\begin{equation}
\epsilon_\beta^*=s_\beta^\epsilon\epsilon,
\qquad
\left(\frac{d\kappa_\beta}{dt}\right)_{\rm term}
=
-\epsilon_\beta^* .
  \label{eq:band-drain-target}
\end{equation}
This equation identifies the terminal-drain share assigned to the band.
The complete band-energy equation also contains rapid production, current-band crossing,
and slow effective-gradient evolution.  In the reference implementation the drain target
\(\epsilon_\beta^*\) is realized through the band-aggregate effective gradients with RTAU matching
of Section~\ref{subsec:effective-gradients-srr} below, using the target as the terminal loss to be matched.
The coefficients $s_\beta^\epsilon$ distribute the modeled terminal loss among retained bands.  Conservative inter-band fluxes $\Pi_{\alpha\to\beta}$ belong to a different model class, and the low-pass coefficients $a_\beta$ used to construct filtered observables play a separate role.

For the reference closure, the initial band values of
\(\omega_{L,\beta}\) are tied to the initial large-scale-enstrophy
distribution.  Let
\begin{equation}
  s_\beta^Z
  =
  \frac{Z_\beta(0)}
       {\sum_\gamma Z_\gamma(0)}
  \label{eq:initial-Z-share}
\end{equation}
denote the initial LSE-moment share of band \(\beta\).  The initial
\(\omega_{L,\beta}\) values are chosen so that the raw active-LSE drain
shares match these \(Z_\beta\) shares.  Equivalently, if
\(\widetilde\epsilon_\beta^{(1)}\) denotes the raw band drain evaluated with
unit \(\omega_{L,\beta}\), then
\begin{equation}
  \omega_{L,\beta}(0)
  =
  \frac{s_\beta^Z\,\epsilon_0}
       {\widetilde\epsilon_\beta^{(1)}} .
  \label{eq:initial-omegaL-band}
\end{equation}
This fixes the initial distribution of the active terminal-drain map
from the retained spectral moments rather than from a case-dependent
tuning choice.

\subsection{Equivalent effective-gradient and SRR equations}
\label{subsec:effective-gradients-srr}

The band-aggregate drain targets are used to set the IPRM effective-gradient time scales.  In the packet equations the effective gradients enter as additions to the mean gradient,
\begin{equation}
  G_{ij}^{v,\beta}=G_{ij}+A_{ij}^{v,\beta},
  \qquad
  G_{ij}^{n,\beta}=G_{ij}+A_{ij}^{n,\beta},
  \label{eq:effective-Gv-Gn-beta}
\end{equation}
with the usual frame-rotation term added to the velocity equation in rotating-frame cases.  Consequently the orientation motion, covariance evolution, and radial drift are all modified by the slow structural response.  For example, the current-wavenumber shift of a packet in band $\beta$ contains
\begin{equation}
  \dot s^{(p)}
  =-n_i^{(p)}G_{ij}^{n,\beta}n_j^{(p)},
  \label{eq:slow-radial-shift}
\end{equation}
so effective gradients can change the migration and residency of packets in current bands.  Their effect is therefore dynamical: they alter the detailed ray--band evolution before the one-point reconstruction is formed.

Following the original IPRM matching logic, we write the effective-gradient drain associated with band $\beta$ as
\begin{equation}
  \epsilon^{\rm PRM}_\beta
  =q_\beta^2\frac{C^v}{\tau_\beta}
   r_{ik}^{\beta}d_{km}^{\beta}r_{mi}^{\beta}
  =q_\beta^2\frac{C^v}{\tau_\beta}
   \operatorname{tr}(r^\beta d^\beta r^\beta).
  \label{eq:epsilon-prm-beta-reference}
\end{equation}
The contraction in Eq.~\eqref{eq:epsilon-prm-beta-reference} is the inherited IPRM structural factor entering the effective-gradient energy accounting.  It is kept as a contraction of the normalized band tensors, and the notation avoids assigning it the role of a new scale variable.  For active, non-degenerate bands, the reference implementation chooses the reciprocal time scale by requiring
\begin{equation}
  \epsilon^{\rm PRM}_\beta=\epsilon_\beta^*,
  \qquad
  \frac{1}{\tau_\beta}
  =\frac{\epsilon_\beta^*}
        {q_\beta^2 C^v\operatorname{tr}(r^\beta d^\beta r^\beta)},
  \label{eq:tau-beta-reference}
\end{equation}
and applies
\begin{equation}
  A_{ij}^{v,\beta}
  =\frac{C^v}{\tau_\beta}r_{ik}^{\beta}d_{kj}^{\beta},
  \qquad
  A_{ij}^{n,\beta}
  =\frac{C^n}{\tau_\beta}r_{ik}^{\beta}d_{kj}^{\beta}.
  \label{eq:AvAn-beta-reference}
\end{equation}
The reciprocal time scale $1/\tau_\beta$ is the band-local, band-aggregate version of the IPRM RTAU matching~\citep{KassinosReynolds1996IPRM}.  It is a high-Re structural time scale chosen so that the inherited effective-gradient dissipation accounting reproduces the terminal drain supplied externally by the active LSE branch.  In the original IPRM this matching was performed globally against the modified $\epsilon$ equation; in the present reference RC-IPRM it is performed separately for each band against the LSE terminal-drain shares while preserving the prescribed global terminal loss.
Thus the effective-gradient dissipation accounting satisfies
\begin{equation}
  \epsilon_{\rm eff}
  =\sum_\beta \epsilon^{\rm PRM}_\beta
  =\sum_\beta \epsilon_\beta^*
  =\epsilon,
  \label{eq:epsilon-eff-matched-reference}
\end{equation}
apart from numerical safeguards in nearly depleted or structurally degenerate bands.  This equality is the reason the LSE shares can be used while preserving the global target drain.

Slow rotational randomization uses the same band-aggregate structure.  Its role is distinct from terminal drain.  SRR represents the slow nonlinear rotational scrambling of the packet covariance within the plane transverse to the packet orientation.  It preserves the packet trace and therefore redistributes componentality while leaving the packet energy unchanged.  The structural rotation available to this process is measured by
\begin{equation}
  \Omega_i^{*,\beta}
  =\epsilon_{ipq}\ r^{\beta}_{qz} d^{\beta}_{zp},
  \qquad
  \omega_*^\beta=\left(\Omega_i^{*,\beta}\Omega_i^{*,\beta}\right)^{1/2}.
  \label{eq:omega-star-beta}
\end{equation}
This quantity vanishes when the band-aggregate componentality and dimensionality tensors are coaxial, and becomes active when their product contains an antisymmetric rotational part.
For a packet $p$ belonging to band $\beta(p)$, the scalar randomization rate used in the reference implementation is
\begin{equation}
  C_r^{(p)}
  =C_{\rm SRR}\frac{1}{\tau_{\beta(p)}}
  \omega_*^{\beta(p)}
  \pospart{f_{ij}^{\beta(p)}n_i^{(p)}n_j^{(p)}}.
  \label{eq:srr-rate-reference}
\end{equation}
The SRR operator itself is trace preserving; the rate in Eq.~\eqref{eq:srr-rate-reference} determines how rapidly the packet covariance is randomized within the plane transverse to its current orientation.  The factor $\pospart{f_{ij}^{\beta}n_i^{(p)}n_j^{(p)}}$ restricts the randomization rate according to the local circulicity seen by that packet direction, making SRR a structural randomization process distinct from isotropic eddy-viscosity damping.

\section{Canonical homogeneous-flow validation}
\label{sec:validation}

The validation set tests the coherence of one reference closure across distinct homogeneous deformations.  The strain cases expose the sensitivity of the active LSE terminal-drain map; homogeneous shear tests the simultaneous evolution of componentality, dimensionality, circulicity, and scalar time scales; the elliptic and rotating-shear cases test the interaction of deformation with mean and frame rotation.  The main figures show representative cases that carry the argument.  The appendix Atlas contains tensor and scalar comparisons for the full irrotational-strain family from the same calculation set; see Appendix~\ref{app:atlas}.

All main validation calculations use the same Ray--Column projection
structure: the same orientation-ray quadrature, the same initial radial
packet density, and the same four current-\(k\) bands.  The initial
spectrum is also the same across the strain, shear, elliptic-streamline,
and rotating-shear cases.  Differences among the figures therefore arise
from the imposed deformation history and from the reference closure
response, not from case-dependent changes in the projection density or
initial spectral discretization.

The numerical implementation used for these calculations is written in
MLX~\citep{MLX_GitHub} and executed on Apple Silicon GPUs.  This computational choice is
only the realization used to advance the ray--packet quadrature; it is
not part of the model definition.  The important modeling point is that
the same finite ray--packet representation is used throughout the main
validation set.

\subsection{Case definitions and initial scale normalizations}
\label{subsec:case-definitions}

The case parameters and initial scale normalizations used in the
validation figures are summarized in Table~\ref{tab:case-definitions}.
The model calculations use the common normalization
\[
  \kappa_0=\frac12,\qquad q_0^2=2\kappa_0=1.
\]
The scale ratios reported in the DNS/LES references therefore determine
the model value of \(\epsilon_0\) once the mean deformation rate normalization for each
case has been fixed.  The entries in the last column of
Table~\ref{tab:case-definitions} are these implied model-unit values,
not fitted parameters.

The deformation definitions are stated before the table to keep the
case ledger compact.  For the irrotational strain cases, \(S=1\) in
model units and the plotted coordinate is $C=\exp(St)$.
The axisymmetric contraction cases AXK and AXL use:
$$
  G/S=\mathrm{diag}(1,-1/2,-1/2),
$$
the axisymmetric expansion cases EXO and EXP use
\[
  G/S=\mathrm{diag}(-1,1/2,1/2),
\]
and the plane-strain cases PXA and PXD use
\[
  G/S=\mathrm{diag}(0,-1,1).
\]
For homogeneous shear, \(G_{12}=S\).  For elliptic-streamline cases,
\[
  G_{13}=-(\gamma+e),\qquad G_{31}=\gamma-e,
\]
with
\[
  E=\left(\frac{\gamma+e}{\gamma-e}\right)^{1/2},
  \qquad
  \gamma=1,
  \qquad
  e=\frac{E^2-1}{E^2+1},
\]
and the plotted coordinate is \(et\).  For rotating homogeneous shear,
\(G_{12}=S\) and the frame rotation is specified by \(\omega^f/S\).

\begin{table}[!htbp]
\centering
\small
\setlength{\tabcolsep}{4pt}
\renewcommand{\arraystretch}{1.12}
\caption{Case definitions and initial scale normalizations used in the
validation figures.  The model normalization is
\(\kappa_0=1/2\) and \(q_0^2=1\).}
\label{tab:case-definitions}
\begin{tabular}{@{}llclc@{}}
\toprule
Case & Flow class & Coord. & Reported scale ratio & \(\epsilon_0\) \\
\midrule
AXK &
axis. contraction &
\(C\) &
\(S q_0^2/\epsilon_0=1.1\) &
\(0.9091\) \\

AXL &
axis. contraction &
\(C\) &
\(S q_0^2/\epsilon_0=11.1\) &
\(0.09009\) \\

EXO &
axis. expansion &
\(C\) &
\(S q_0^2/\epsilon_0=0.82\) &
\(1.2195\) \\

EXP &
axis. expansion &
\(C\) &
\(S q_0^2/\epsilon_0=8.2\) &
\(0.12195\) \\

PXA &
plane strain &
\(C\) &
\(S q_0^2/\epsilon_0=1.0\) &
\(1.0000\) \\

PXD &
plane strain &
\(C\) &
\(S q_0^2/\epsilon_0=8.0\) &
\(0.1250\) \\

HS &
hom. shear &
\(St\) &
\begin{tabular}[t]{@{}l@{}}
\(S\kappa_0/\epsilon_0=2.36\)\\
\((S q_0^2/\epsilon_0=4.72)\)
\end{tabular}
&
\(0.21186\) \\

e2 &
elliptic, \(E=1.25\) &
\(et\) &
\begin{tabular}[t]{@{}l@{}}
\(\gamma\kappa_0/\epsilon_0=7.68481\)\\
\((e\kappa_0/\epsilon_0=1.68691)\)
\end{tabular}
&
\(0.06506\) \\

e4 &
elliptic, \(E=2.0\) &
\(et\) &
\begin{tabular}[t]{@{}l@{}}
\(\gamma\kappa_0/\epsilon_0=2.81152\)\\
\((e\kappa_0/\epsilon_0=1.68691)\)
\end{tabular}
&
\(0.17784\) \\

Bardina RSH &
rotating hom. shear &
\(St\) &
\begin{tabular}[t]{@{}l@{}}
\(S q_0^2/\epsilon_0=4.036\)\\
\(\omega^f/S=-1,0,0.5,1\)
\end{tabular}
&
\(0.24777\) \\
\bottomrule
\end{tabular}
\end{table}
\break

\subsection{Irrotational strain}
\label{subsec:strain-validation}

\begin{figure}[!htbp]
  \centering
  \includegraphics[width=0.80\textwidth]{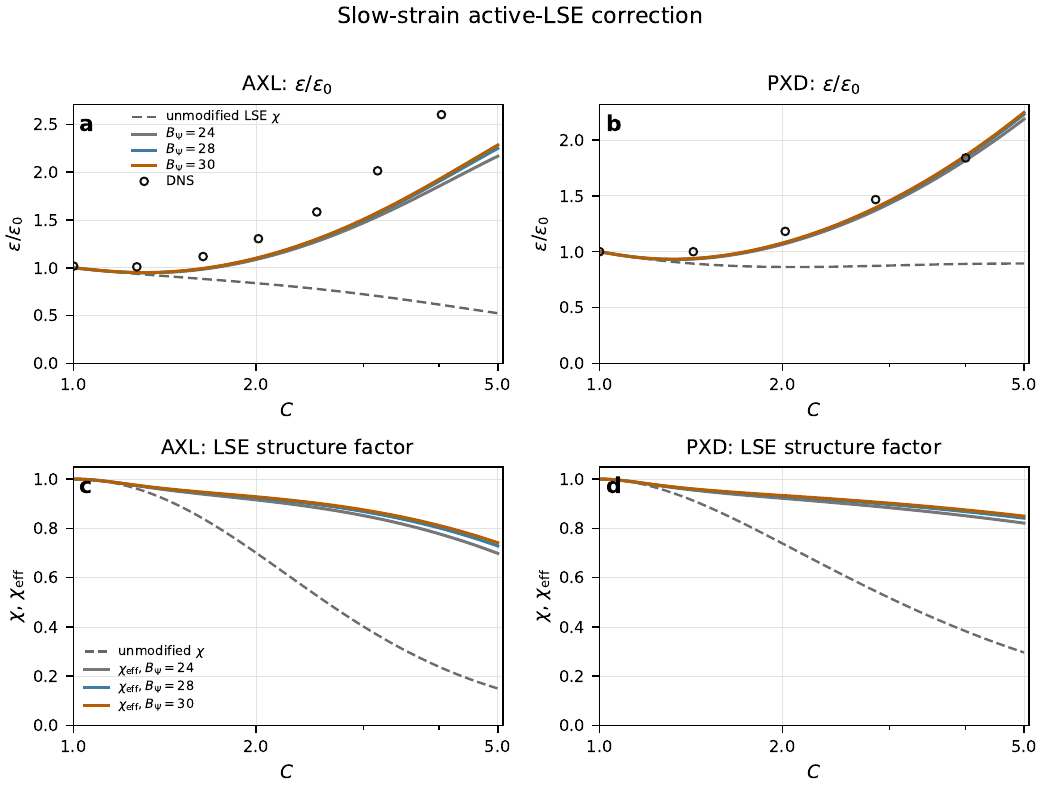}
  \caption{Slow-strain active-LSE correction.  Panels (a,b) compare normalized terminal drain $\epsilon/\epsilon_0$ for representative difficult strain cases; open circles denote DNS data from Lee and Reynolds~\citep{LeeReynolds1985}, dashed gray curves denote the unmodified active-LSE map, and colored curves show the corrected map for the indicated $B_\Psi$ values.  Panels (c,d) show the corresponding raw and corrected LSE structure factors.  The unmodified factor $\chi=3f:d$ can collapse in slow strain, while the $\Psifd$ correction regularizes the active LSE structure-to-dissipation map through a continuous invariant constructed from the deviatoric circulicity and dimensionality tensors.  The correction applies to active use of the LSE map of \citet{ReynoldsLangerKassinos2002}; legacy IPRM used a modified $\epsilon$ equation for the second scale.}
  \label{fig:lse-correction-strain}
\end{figure}

The irrotational strain family provides the sharpest test of the active
LSE correction.  In these cases the unmodified LSE map can continue to
evolve the large-scale enstrophy variable \(\omega_L\), while the
structural transfer factor
$ \chi = 3 f_{ij}d_{ji}$
collapses as the dimensionality and circulicity tensors become
increasingly anisotropic and complementary.  The resulting failure mode
is most evident for intermediate strain-rate cases.  Figure~\ref{fig:lse-correction-strain}
therefore emphasizes AXL and PXD as representative difficult examples.
The comparison should be read as a test of mechanism, with optimized curve fitting left outside the intended interpretation.  The \(f\)--\(d\) complementarity factor restores a physically
reasonable terminal-drain map in slow-strain states where dimensionality
and circulicity remain strongly structured but poorly aligned for
terminal transfer.  At the same time, the correction preserves the
isotropic and 2D--2C limits discussed above.  The same representative
value of \(B_\Psi\) is then retained in the shear, elliptic, and
rotating-shear calculations presented below and in the irrotational-strain atlas
reported in Appendix~\ref{app:atlas}.

\subsection{Homogeneous shear}
\label{subsec:hs-validation}
\begin{figure}[!htbp]
  \centering
  \includegraphics[width=0.80\textwidth]{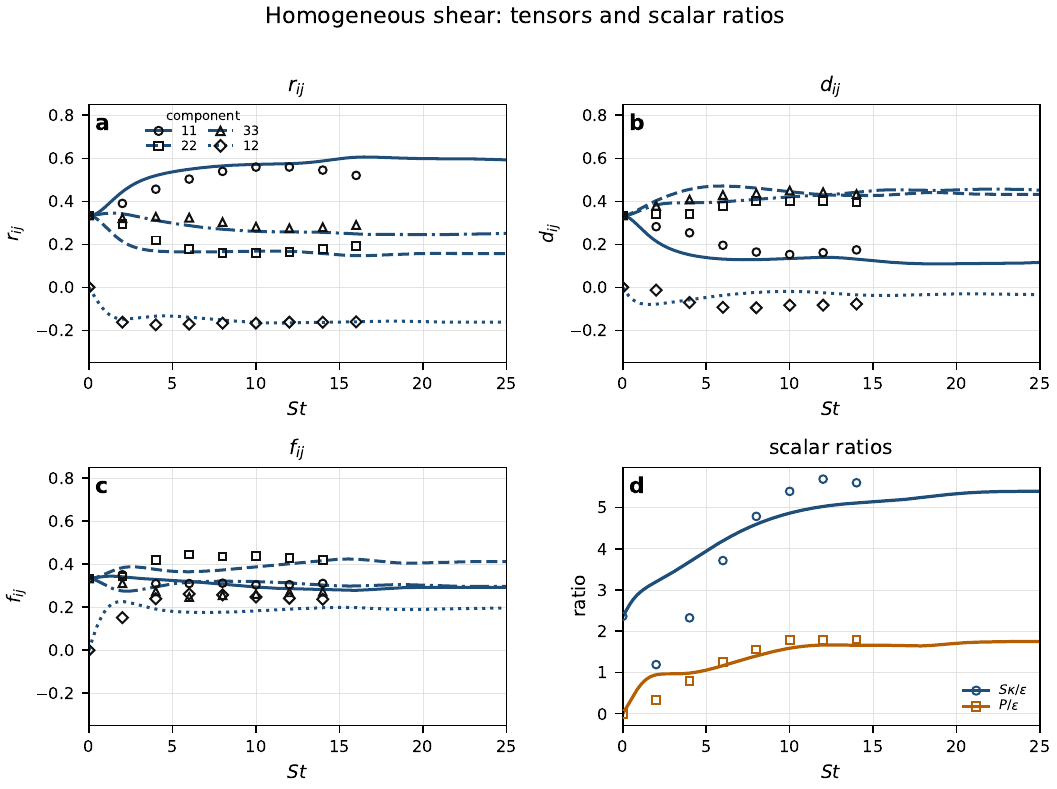}
  \caption{Homogeneous-shear validation.  Lines denote the reference RC-IPRM closure and open markers denote homogeneous-shear data~\citep{RogersMoinReynolds1986}; component identity is encoded by line style and marker shape.  The same closure constants used in the strain cases are retained, and the horizontal axis is restricted to $St\le 25$ to emphasize the approach to the developed regime while keeping the early transient readable.  The tensor panels test componentality, dimensionality, and circulicity simultaneously, while panel (d) compares the scalar ratios $S\kappa/\epsilon$ and $P/\epsilon$.  The purpose is to verify that the active-LSE correction and band-aggregate closure preserve the stress and structure evolution while correcting selected scalar behavior.}
  \label{fig:hs-validation}
\end{figure}

Homogeneous shear provides the main quantitative backbone case outside
irrotational strain.  It combines sustained production, slow structural
redistribution, and terminal drain in a single deformation history, and
therefore tests whether the closure that repairs the scalar drain in
strain also preserves the coupled tensor dynamics.  In
Fig.~\ref{fig:hs-validation}, the tensor panels compare the normalized
componentality, dimensionality, and circulicity tensors simultaneously,
while the scalar panel compares the production and shear time-scale
ratios, including \(P/\epsilon\) and \(S\kappa/\epsilon\).

The agreement varies by component, but the same reference closure that
regularizes the slow-strain drain gives a coherent homogeneous-shear
response with the same constants and the same band-aggregate closure
level.  In particular, the predicted component levels of all three
normalized one-point structure tensors remain in reasonable agreement
with the comparison data over the shear history.  This indicates that the
band-aggregate evaluation of the slow and terminal ingredients preserves a
credible approximation of the evolving turbulence structure in addition
to correcting the scalar energy balance.

\section{Rotation and filtered-observable guards}
\label{sec:rotation-guards}

\subsection{Elliptic-streamline cases}
\label{subsec:elliptic}

The elliptic-streamline comparisons are used as qualitative
rotation/instability guards, with high-Reynolds-number calibration left
to the shear and strain comparisons~\citep{BlaisdellShariff1994,BlaisdellShariff1996}.
The DNS data are low-Reynolds-number calculations and, as discussed by
Blaisdell and Shariff, are affected by finite-domain limitations.  The
useful question is whether the present high-Reynolds-number
homogeneous structural model gives the correct qualitative branch of
behavior: suppression versus growth, delayed instability, and the
ordering of the dominant stress components.

\begin{figure}[!htbp]
  \centering
  \includegraphics[width=0.80\textwidth]{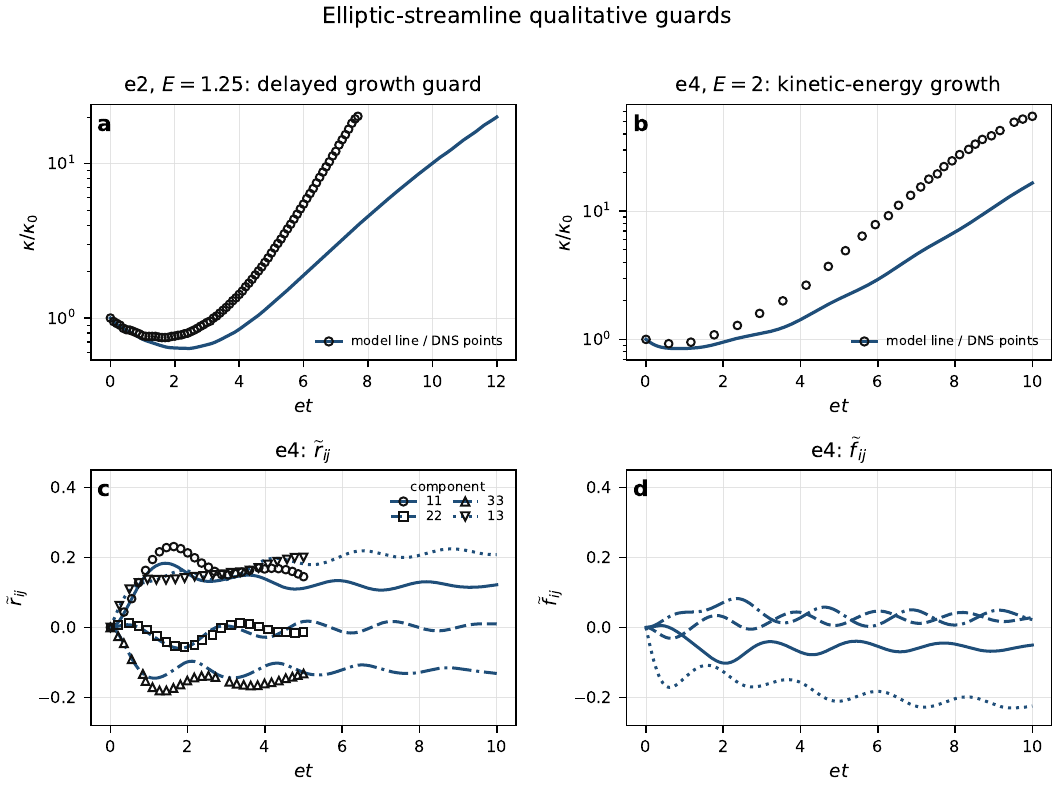}
  \caption{Elliptic-streamline qualitative guards.  Lines denote the reference RC-IPRM closure and open markers denote DNS data from Blaisdell and Shariff~\citep{BlaisdellShariff1994,BlaisdellShariff1996}.  The e2 and e4 cases test whether the closure responds plausibly to elliptic rotation and instability; the high-Reynolds-number calibration role is carried by the strain and shear cases.  Panels (a,b) guard against qualitatively wrong energy behavior, such as immediate decay where delayed growth is expected, while panels (c,d) check the stress-structure branch ordering in the adopted elliptic convention.  Energy is reported as $\kappa/\kappa_0$.}
  \label{fig:elliptic-guard}
\end{figure}

Figure~\ref{fig:elliptic-guard} is included for this purpose.  The case
labeled \(1.25\) is especially important because it lies in a regime
where many one-point closures predict decay of the turbulent kinetic
energy, whereas the DNS exhibits growth after the instability develops.
The same closure constants and projection settings are used for the
elliptic cases.  They act as guards against a scale-conditioned closure
that behaves plausibly in irrotational strain and homogeneous shear but
fails under combined strain and rotation. In this sense the elliptic
cases test whether the band-aggregate closure preserves the
instability-sensitive structure of the original PRM/IPRM dynamics in
addition to improving scalar drain behavior.

\subsection{Rotating homogeneous shear and Bardina filtered LES}
\label{subsec:bardina}

\begin{figure}[!htbp]
  \centering
  \includegraphics[width=0.80\textwidth]{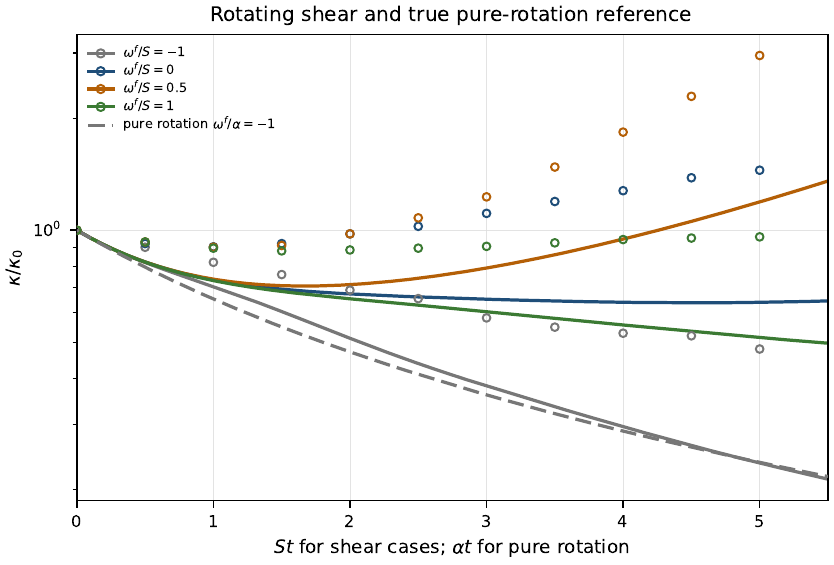}
  \caption{Bardina rotating shear: global energy observable.  Lines show global $\kappa/\kappa_0$ from the reference Ray--Column closure and open markers show filtered LES data from Bardina et al.~\citep{BardinaFerzigerReynolds1983}, with color denoting the frame-vorticity ratio $\omega^f/S$.  The dashed gray curve is a true pure-rotation reference with $G=0$ and $\omega^f/\alpha=-1$, where $\alpha$ is the imposed pure-rotation rate used for that reference; it is included to separate the pure-rotation behavior from the sheared $\omega^f/S=-1$ case.  Because the comparison data are filtered LES quantities, this global observable is interpreted as a useful diagnostic.  The filtered low-pass observable in Fig.~\ref{fig:bardina-lowpass} is the more direct use of the retained Ray--Column band structure.}
  \label{fig:bardina-global}
\end{figure}

\begin{figure}[!htbp]
  \centering
  \includegraphics[width=0.80\textwidth]{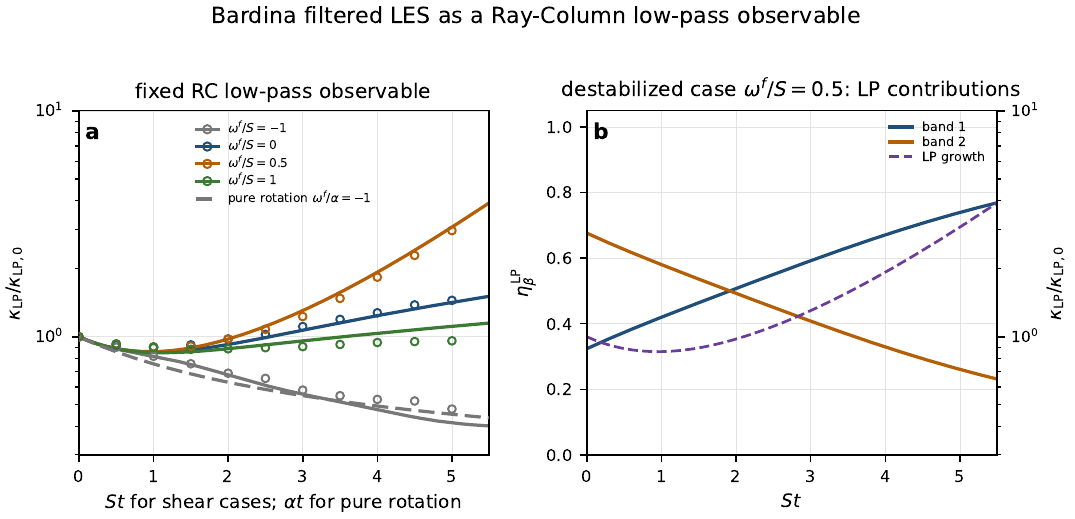}
  \caption{Bardina rotating shear: fixed Ray--Column low-pass observable.  Because the Bardina data are filtered LES quantities, the retained band energies can be combined before one-point averaging to form $\kappa_{\rm LP}=\sum_\beta a_\beta\kappa_\beta$.  Panel (a) compares this low-pass energy growth with the Bardina filtered LES points~\citep{BardinaFerzigerReynolds1983} using the same low-pass operator for all rotating-shear cases; lines are RC-IPRM low-pass observables and open markers are LES data.  Panel (b) shows the effective contribution fractions $\eta_\beta^{\rm LP}=a_\beta\kappa_\beta/\sum_\gamma a_\gamma\kappa_\gamma$ for the destabilized $\omega^f/S=0.5$ case, together with the low-pass growth on the right axis.  This figure is the main demonstration that retained radial bands carry observable information: they allow filtered observables to be constructed before scale information is lost.  The coefficients $a_\beta$ are projection-specific filter coefficients and are kept separate from universal closure constants.}
  \label{fig:bardina-lowpass}
\end{figure}

For homogeneous shear in a rotating frame, referred to below as rotating
shear, the comparison is parameterized by the frame-vorticity ratio
\(\omega^f/S\).  Bardina's rotating-shear data are filtered LES
quantities, distinct from unfiltered DNS statistics~\citep{BardinaFerzigerReynolds1983}.
This distinction is important.  A global one-point energy comparison is
still informative because it tests the rotating-shear trend.  The
observable most directly matched to the LES data is the corresponding
low-pass comparison formed from the retained band populations, shown in
Fig.~\ref{fig:bardina-lowpass}.  Figure~\ref{fig:bardina-global} is
therefore interpreted as a global diagnostic.

The Ray--Column representation allows an additional fixed low-pass observable,
\begin{equation}
  \kLP(t)=\sum_\beta a_\beta\kappa_\beta(t),
  \qquad
  \frac{\kLP(t)}{\kLP(0)}.
  \label{eq:lowpass-observable}
\end{equation}
The same low-pass operator is applied to all Bardina rotating-shear cases.  The coefficients $a_\beta$ are tied to the current finite band projection and to the chosen filtered observable, with the same values used across the frame-rotation ratios and kept separate from the unfiltered reconstruction.  Figure~\ref{fig:bardina-lowpass} is consequently the most direct use of the retained radial bands in the present validation set.  It shows that band information can be used to construct a filtered observable before the one-point average has destroyed the radial distribution of energy.  This is a stronger demonstration of the Ray--Column idea than the global energy curve alone.

Taken together, Figs.~\ref{fig:bardina-global} and~\ref{fig:bardina-lowpass} separate two claims.  The global curves show that the reference closure responds plausibly to frame rotation in homogeneous shear.  The low-pass curves show something more specific to RC-IPRM: retained band energies can be recombined into a filtered observable using the same closure at the comparison stage.  The result demonstrates why scale conditioning is useful even in a reduced homogeneous particle representation.

\subsection{Realizability audit}
\label{subsec:realizability-audit}

\begin{figure}[!htbp]
  \centering
  \includegraphics[width=0.98\textwidth]{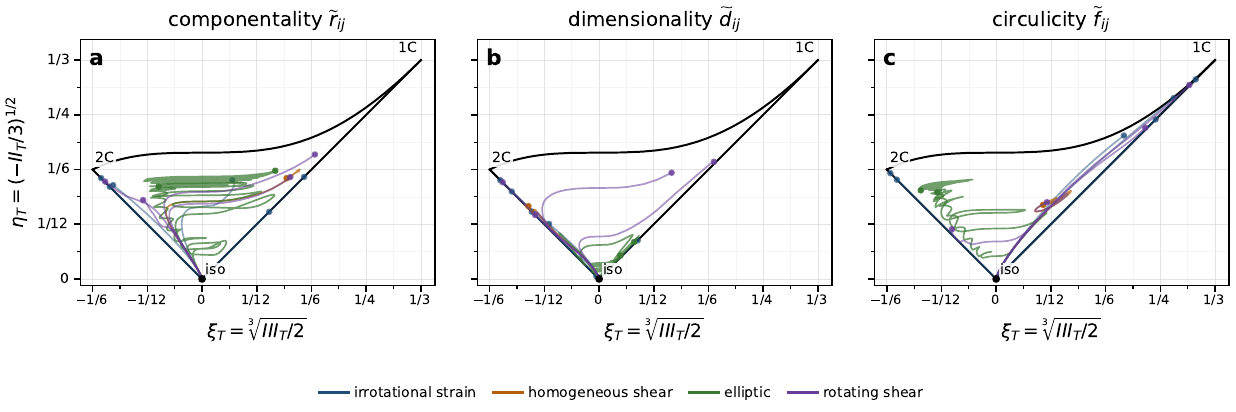}
  \caption{Lumley invariant-map audit for the normalized structure
  tensors across the validation histories.  The three panels show the
  deviatoric componentality tensor \(\widetilde r_{ij}\), dimensionality
  tensor \(\widetilde d_{ij}\), and circulicity tensor
  \(\widetilde f_{ij}\).  Trajectories are grouped by flow family:
  irrotational strain, homogeneous shear, elliptic-streamline flow, and
  rotating shear.  The plotted states include the main-body validation
  cases and the irrotational-strain Atlas cases.  No sampled trajectory
  leaves the Lumley realizability domain; the figure is a numerical audit
  of the computed cases rather than a formal realizability proof.}
  \label{fig:lumley-realizability-audit}
\end{figure}

As a final check on the computed validation histories, we examine the
normalized structure tensors on the Lumley invariant map~\citep{LumleyNewman1977}.  For
\(T_{ij}\in\{r_{ij},d_{ij},f_{ij}\}\), let
\[
  \widetilde T_{ij}=T_{ij}-\frac13\delta_{ij},
\]
and define
\[
  II_T=-\frac12\widetilde T_{ij}\widetilde T_{ji},
  \qquad
  III_T=\frac13\widetilde T_{ij}\widetilde T_{jk}\widetilde T_{ki}.
\]
The plotted coordinates are
\[
  \eta_T=\left(-\frac{II_T}{3}\right)^{1/2},
  \qquad
  \xi_T=\sqrt[3]{\frac{III_T}{2}},
\]
where the cube root is the real cube root.  Figure~\ref{fig:lumley-realizability-audit} shows the trajectories of the
componentality, dimensionality, and circulicity tensors for all cases
considered in the main validation set and for the irrotational-strain
Atlas cases reported in Figs.~\ref{fig:atlas-AXK}--\ref{fig:atlas-PXD}
of Appendix~\ref{app:atlas}.  No sampled state leaves the realizability
domain.  This is a numerical audit of the computed trajectories,
complementing the packet-based construction of the structure tensors,
rather than a formal proof of strong realizability for the continuous
closure.  A formal proof would require showing that the governing
equations satisfy the appropriate boundary-derivative conditions as
eigenvalues of the structure tensors approach zero.

Overall, the comparisons in this section and in Appendix~\ref{app:atlas} indicate modest but useful improvements in several
tensorial and scalar quantities relative to the original IPRM.  The more important outcome, however,
is not a large empirical reduction of error relative to the parent
structure-based IPRM model.  The main advance is that the Ray--Column
formulation restores a reduced radial spectral representation while
preserving the PRM/IPRM interpretation of componentality,
dimensionality, and circulicity.  This makes the slow structural
response, terminal-drain assignment, and filtered observables part of a
single retained structure--scale state.

\section{Interpretation, limitations, and model parsimony}
\label{sec:limitations}

\subsection{Meaning of the bands}
\label{subsec:band-meaning}

Finite Ray--Column bands are scale-conditioned spectral-structural
populations.  They delay global radial averaging until the effective-gradient,
slow-randomization, and terminal-drain coefficients have been evaluated from
orientation- and wavenumber-integrated band structure.  In this sense the bands
define the structural populations on which the reference closure is evaluated.

The retained bands serve as projection intervals for the structural
populations used by the closure.  Conservative nonlinear inter-band flux
models of the form \(\Pi_{\alpha\to\beta}\), with zero net transfer
over the modeled bands, belong to a later model class.  Current-band
crossing under rapid distortion is a kinematic consequence of wavevector
evolution and is handled by the packet projection described in
Appendix~\ref{app:rdt-consistency}.  The active LSE branch supplies a
terminal-drain assignment for the retained bands.

\subsection{Finite projection and finite-domain effects}
\label{subsec:finite-box}

The Ray--Column formulation represents a spectral-structural ensemble
while DNS and LES simulate periodic velocity fields.  Its numerical limitations are
those of a finite projection: finite orientation quadrature, finite
packet sampling, finite radial resolution, and reduced reliability in
bands that become weakly populated or energetically depleted.  These are
different from the finite-box effects that can occur in DNS or LES.

This distinction is important when comparing with DNS or LES data.  A
finite-domain DNS limitation reflects the physical and numerical
constraints of the simulated velocity field.  A Ray--Column limitation
reflects the accuracy with which the ray--packet ensemble and the associated band-aggregate structural populations
are represented and projected.  They are different sources of uncertainty,
even when they appear in the same comparison figure.

\subsection{High-Re terminal drain}
\label{subsec:terminal-drain}

The active LSE closure supplies a high-Reynolds-number terminal loss
rate from the modeled energy-containing structural field.  The present
reference closure represents this loss through a scale-conditioned
terminal-drain model; dissipative-range resolution, conservative cascade
dynamics, and final viscous conversion to heat remain outside the model.

In the reference implementation the band-aggregate LSE map assigns the
target drain shares \(\epsilon_\beta^*\), and the matched IPRM time
scales are chosen so that the effective-gradient drain accounting
realizes those shares.
Thus the LSE branch determines how the high-Re terminal loss is
distributed across bands, while the structural IPRM machinery determines
how that loss is expressed through the band-aggregate componentality,
dimensionality, and circulicity tensors.

\subsection{Role of the effective gradients}
\label{subsec:effective-gradient-role}

The Ray--Column representation also clarifies the dynamical role of the
effective gradients.  In the original one-point presentation they are
easy to view mainly through their energy accounting, because the RTAU
matching sets a time scale that reproduces the externally supplied
terminal drain.  In the band-resolved representation their broader role
is more visible.  The tensors \(A_{ij}^{v,\beta}\) and
\(A_{ij}^{n,\beta}\) in Eq.~\eqref{eq:AvAn-beta-reference} modify the
gradients experienced by the packet velocity covariance and by the packet
orientation.  They therefore alter the packet trajectories in spectral
space, not only the scalar rate at which energy is removed.

This effect appears directly in the current-wavenumber drift.  Since the
radial shift satisfies Eq.~\eqref{eq:slow-radial-shift}, the band-local
effective gradient \(G_{ij}^{n,\beta}\) changes how packets move across
the current \(k\)-bands.  The effective gradients therefore influence
band residency, band migration, and the sequence of structures that enter
the band-aggregate tensors used by the slow and terminal closures.  In
this sense the terminal-drain matching fixes the energetic scale of the
slow response, while the effective-gradient tensors determine how that
response acts on the structural packet ensemble.

This distinction is one reason the same reference closure can be tested
across irrotational strain, homogeneous shear, elliptic-streamline flow,
and rotating shear without changing its constants.  The comparisons do
not imply that the effective-gradient construction is unique, but they
show that it supplies a detailed dynamical modification of the PRM/RDT
packet evolution rather than a simple scalar damping of that evolution.

\subsection{Parsimony of the reference closure}
\label{subsec:model-parsimony}

The present comparisons use the same reference RC--IPRM closure across
irrotational strain, homogeneous shear, elliptic-streamline flow, and
rotating shear.  The \(\Psi_{fd}\) correction regularizes a specific
slow-strain failure mode of the active LSE map, and the representative
value of \(B_\Psi\) is then retained in the subsequent cases.

This parsimony is essential to the interpretation of the results.  A
single scale-conditioned structural closure preserves the exact
rapid/RDT limit, recovers the global IPRM limit when the slow closures
are evaluated globally, and gives coherent behavior across several
distinct homogeneous deformation classes when the slow and terminal
ingredients are evaluated band locally but from band-aggregate tensors.
The result should be read as a coherent reference model, with componentwise optimal tuning for each flow left outside the present claim.

The numerical projection was also kept fixed across the main validation
figures.  The same orientation-ray density, radial band structure, and
radial \(k\)-packet quadrature were used for irrotational strain,
homogeneous shear, elliptic-streamline flow, and rotating shear.  Any
remaining projection error is therefore part of a common numerical
representation, separate from case-dependent tuning.

\subsection{Representative constants}
\label{subsec:constants}

The reference closure is presented with a small number of physical constants as summarized in Table~\ref{tab:constants}.  Numerical discretization choices, such as the number of rays, packets, or bands, are kept separate from physical closure constants.

\begin{table}[!htbp]
\centering
\caption{Physical constants and representative settings in the current reference closure. Numerical quadrature and band-discretization choices are reported separately from physical closure constants.}
\label{tab:constants}
\begin{tabular}{lll}
\toprule
Quantity & Value & Role \\
\midrule
$\CSRR$ & 8.5 & slow rotational randomization \\
$\Bpsi$ & 28--30; 30 used & complementarity-map strength \\
$C_T^*$ & $3/2$ & LSE $k^4$ spectrum transfer coefficient \\
$C_P^*$ & $4/5$ & LSE $k^4$ spectrum production coefficient \\
$C^v$ & 1 & effective-gradient velocity-scale normalization \\
$C^n/C^v$ & 2.2 & effective-gradient orientation/velocity ratio \\
\bottomrule
\end{tabular}
\end{table}

\section{Discussion and future modeling directions}
\label{sec:discussion-future}

The scale-conditioned PRM framework opens several model paths beyond the reference closure used here.  These include closures closer to the original PRM/IPRM with global terminal drain, alternative band-share rules, cross-band structural randomization, and extensions with conservative inter-band transfer and an explicit dissipative reservoir.  The main validation claim of the present paper concerns the minimal coherent reference model: PRM ray--packet structure, radial spectral-scale retention, band-aggregate active LSE terminal drain, and the $\Psifd$ correction to the active LSE structure-to-dissipation map.

The ability to form low-pass observables from retained band populations
also suggests possible future connections with reduced LES-like
closures; such extensions are left for later work.

Thus, the validation figures should be interpreted with this perspective.  The
reference RC--IPRM closure improves several scalar and tensorial
comparisons, but the improvements are generally moderate rather than
dramatic.  This is not surprising: the parent IPRM already provides a
credible high-Reynolds-number description of many homogeneous
deformation histories.  The principal gain is about the structure of the RC-IPRM.  The model now
retains a finite radial spectral representation, preserves the exact
rapid/RDT limit, evaluates the slow and terminal ingredients on
band-aggregate structural populations, and can form filtered observables
before the one-point average is taken. 

An important part of this structure--scale consistency is the use of a
native structure-based second-scale equation.  The active LSE equation
provides the terminal-drain map from the same componentality,
dimensionality, and circulicity tensors that define the PRM/IPRM
structural state, and its band-local form assigns the terminal drain
across the retained radial spectral populations.

The main physical message is that the Ray--Column representation makes
the scale dependence of the structural response explicit.  The retained
bands show how componentality, dimensionality, and circulicity are
distributed among energy-containing scale populations before the
one-point average is formed.  Within this retained structure--scale
state, the effective gradients act as spectral-trajectory modifiers, the
active LSE equation supplies a native structure-based second-scale map,
and filtered observables can be assembled directly from band
populations.

\section{Conclusions}
\label{sec:conclusions}

The PRM/IPRM conditional representation can be lifted from $R_{ij}^{\vert n}$ to $R_{ij}^{\vert n,k}$, restoring radial spectral scale to structure-based turbulence modeling while preserving the original orientation-conditioned logic. Finite Ray--Column bands are projections of this continuous conditional representation and provide scale-conditioned structural populations for closure evaluation. The implementation realizes the projected theory through ray--packet ensemble sums with the relevant quadrature measures carried by the packets, while the nonlinear slow and terminal closure coefficients are formed from band-aggregate structural tensors. The switch from the legacy modified-$\epsilon$ equation to active LSE exposes a structural failure of the unmodified $\chi=3f:d$ map in slow-strain states; the continuous complementarity invariant $\Psifd$ regularizes that active LSE map. The regenerated validation calculations with this invariant form give coherent behavior across the current main cases and enable filtered or scale-conditioned observables, including the low-pass Bardina comparison. Future work should add conservative inter-band transfer and an explicit dissipative reservoir.

The contribution of RC--IPRM is therefore not best measured by a large
case-by-case reduction of scalar error.  Its value is the construction
of a self-consistent structure--scale closure that retains enough radial
spectral information to support band-local slow response, active LSE
terminal-drain assignment, and filtered-observable construction within
the PRM/IPRM ensemble framework. Compared with the modified \(\epsilon\)-equation used in the original
IPRM calculations, the active LSE formulation gives RC--IPRM a
structure-based second-scale equation whose variables are defined within
the same componentality--dimensionality--circulicity framework as the
rest of the model.

\section*{Data Availability}
The data that support the findings of this study are available from the corresponding author upon reasonable request.

\section*{Acknowledgments}

I gratefully acknowledge the late Prof. William C. Reynolds, my PhD advisor and collaborator.  The PRM/IPRM and structure-tensor line of work grew out of my doctoral research in his group and from the exceptionally creative collaboration that followed.  His physical insight, criticism, and intellectual generosity shaped that period of work in ways that continue to influence how I think about turbulence.

\appendix
\section{RDT consistency of the Ray--Column projection}
\label{app:rdt-consistency}

This appendix records the rapid-distortion consistency of the finite-band
projection.  The radial bands used in RC--IPRM are projections of the
same PRM/RDT spectral evolution, with the rapid dynamics inherited from
the parent model.

Let
\begin{equation}
  \xi=\log k,
  \qquad
  k=|\bm N|,
  \qquad
  \bm N=k n,
\end{equation}
and define \(\xi_\beta=\log k_\beta\) for the band boundary
\(k_\beta\).  The log-radial tensor density is
\begin{equation}
  \mathcal Q_{ij}^{\vert n,\xi}(t)
  =
  k\,\mathcal R_{ij}^{\vert n,k}(t)
  =
  k^3\Phi_{ij}(k n,t).
  \label{eq:log-density-Q}
\end{equation}
Then
\begin{equation}
  R_{ij}^{\vert n,\beta}(t)
  =
  \int_{\xi_{\beta-1}}^{\xi_\beta}
  \mathcal Q_{ij}^{\vert n,\xi}(t)\,\mathrm d\xi .
  \label{eq:log-band-R}
\end{equation}

For homogeneous RDT the spectral vector evolves as
\begin{equation}
  \dot N_i=-G_{ji}N_j,
  \label{eq:app-Ndot}
\end{equation}
where \(G_{ij}\) is the mean-velocity gradient.  Hence
\begin{equation}
  \dot\xi
  =
  -G_{ij}n_i n_j
  =
  -S_{ij}n_i n_j,
  \qquad
  S_{ij}=\frac12(G_{ij}+G_{ji}),
  \label{eq:app-xidot}
\end{equation}
and the orientation satisfies
\begin{equation}
  \dot n_i
  =
  -G_{ji}n_j
  +
  G_{mn}n_m n_n n_i .
  \label{eq:app-ndot}
\end{equation}
Let \(D_n/Dt\) denote differentiation following this orientation
characteristic.  The incompressible RDT velocity-amplitude equation can
be written
\begin{equation}
  \dot u_i=M_{ik}u_k,
  \qquad
  M_{ik}=-G_{ik}+2n_i n_\ell G_{\ell k},
  \label{eq:app-Mik}
\end{equation}
where the second term is the pressure projection that preserves
\(u_i n_i=0\).  The corresponding linear stress operator is
\begin{equation}
  \mathcal L_{ij}[G;A,n]
  =
  M_{ik}A_{kj}+M_{jk}A_{ik}.
  \label{eq:app-Lij}
\end{equation}

For a band defined by current wavenumber, the finite-volume form of the
RDT equation is
\begin{equation}
  \frac{D_n R_{ij}^{\vert n,\beta}}{Dt}
  =
  \int_{\xi_{\beta-1}}^{\xi_\beta}
  \mathcal L_{ij}\!\left[
    G;\mathcal Q^{\vert n,\xi},n
  \right]\,\mathrm d\xi
  -
  \left[
    \dot\xi\,\mathcal Q_{ij}^{\vert n,\xi}
  \right]_{\xi_{\beta-1}}^{\xi_\beta}.
  \label{eq:app-band-rdt}
\end{equation}
The last term is the rapid current-band boundary flux.  It exists
because the mean deformation changes the magnitude of the spectral
vector, and it represents kinematic boundary crossing in current
wavenumber.  If
\begin{equation}
  \mathcal F_{ij,\beta+1/2}^{\vert n}
  =
  \dot\xi\,\mathcal Q_{ij}^{\vert n,\xi}(\xi_\beta,n,t),
  \label{eq:app-boundary-flux}
\end{equation}
then the boundary contribution to band \(\beta\) is
\begin{equation}
  \mathcal F_{ij,\beta-1/2}^{\vert n}
  -
  \mathcal F_{ij,\beta+1/2}^{\vert n}.
\end{equation}
Summing over a complete band partition gives
\begin{equation}
  \sum_\beta
  \left(
    \mathcal F_{ij,\beta-1/2}^{\vert n}
    -
    \mathcal F_{ij,\beta+1/2}^{\vert n}
  \right)
  =
  0,
  \label{eq:app-flux-telescoping}
\end{equation}
apart from end-point fluxes.  Thus the internal current-band boundary
fluxes telescope and the ray-integrated PRM/RDT equation is recovered.
Equivalently,
\begin{equation}
  \sum_\beta
  \mathcal L_{ij}\!\left[
    G;R^{\vert n,\beta},n
  \right]
  =
  \mathcal L_{ij}\!\left[
    G;\sum_\beta R^{\vert n,\beta},n
  \right],
  \label{eq:app-L-telescope}
\end{equation}
because \(\mathcal L_{ij}\) is linear in the unnormalized stress tensor.

In the computations reported in this paper, the packet representation
realizes Eq.~\eqref{eq:app-band-rdt} by carrying the ray--packet
quadrature and the evolving radial shift
\begin{equation}
  s^{(p)}(t)
  =
  \log\frac{k^{(p)}(t)}{k_0^{(p)}},
  \qquad
  k^{(p)}(t)=k_0^{(p)}e^{s^{(p)}(t)}.
  \label{eq:app-log-shift}
\end{equation}
In the RDT limit,
\begin{equation}
  \dot s^{(p)}
  =
  -G_{ij}n_i^{(p)}n_j^{(p)} .
  \label{eq:app-sdot}
\end{equation}
A current band \([k_{\beta-1},k_\beta)\) therefore maps, for trajectory
\(p\), to the initial-wavenumber interval
\begin{equation}
  a_{\beta p}=k_{\beta-1}e^{-s^{(p)}},
  \qquad
  b_{\beta p}=k_{\beta}e^{-s^{(p)}}.
  \label{eq:app-current-band-preimage}
\end{equation}
Band moments are then obtained by binning or projecting the current
packet population according to \(k^{(p)}(t)\), as in
Eqs.~\eqref{eq:implemented-Rbeta}--\eqref{eq:implemented-Hbeta}.  In
this representation the boundary-crossing flux is implicit in the
motion of packets across current-band boundaries and is handled by the
same projection used for the band moments.

For the compressed RDT diagnostic implementation, this statement can be
made more explicit.  Let \(p\) label a compressed RDT trajectory and let
\[
  \mathcal I_E^{(0)}(a,b)
  =
  \int_a^b E_0(k_0)\,\mathrm dk_0
\]
denote the initial scalar energy contained in an initial-wavenumber
interval.  If \(C_{ij}^{(p)}\) is the velocity covariance carried by
trajectory \(p\), a current-band stress may be written schematically as
\begin{equation}
  R_{ij}^{\beta}
  =
  \sum_p
  \Omega_p C_{ij}^{(p)}
  \mathcal I_E^{(0)}(a_{\beta p},b_{\beta p}),
  \label{eq:app-compressed-band-stress}
\end{equation}
where \(\Omega_p\) is the fixed quadrature normalization of the trajectory
ensemble.  Differentiating Eq.~\eqref{eq:app-compressed-band-stress}
separates the RDT rate into an amplitude-distortion part and a
radial-crossing part.  The latter has the form
\begin{equation}
  \left(\frac{dR_{ij}^{\beta}}{dt}\right)_{\rm radial}
  =
  \sum_p
  \Omega_p C_{ij}^{(p)}\dot s^{(p)}
  \left[
    a_{\beta p}E_0(a_{\beta p})
    -
    b_{\beta p}E_0(b_{\beta p})
  \right].
  \label{eq:app-radial-crossing-diagnostic}
\end{equation}
This term is a computable diagnostic of rapid current-band crossing.
Bands labeled by the initial wavenumber \(k_0\) provide useful control
partitions with zero current-boundary crossing, whereas current-wavenumber
bands are the relevant objects for low-pass and filtered observables.

\section{Atlas figures for the irrotational-strain family}
\label{app:atlas-planned}
\label{app:atlas}

The main text uses AXL and PXD to discuss the active-LSE correction in two representative difficult irrotational-strain cases.  The corresponding scalar histories are shown in Fig.~\ref{fig:lse-correction-strain}; the present appendix records the tensor and scalar comparisons for the full irrotational-strain family from the same calculation set.  The purpose is limited: the figures document that the adopted RC--IPRM closure, with the same value of \(B_\Psi\), the same projection density, the same initial spectrum, and the same four current-\(k\) bands, gives a coherent response across the strain cases.  Homogeneous shear, elliptic-streamline behavior, and Bardina rotating shear are already represented in the main figures and are therefore not repeated here.

\begin{figure}[!htbp]
  \centering
  \includegraphics[width=0.93\textwidth]{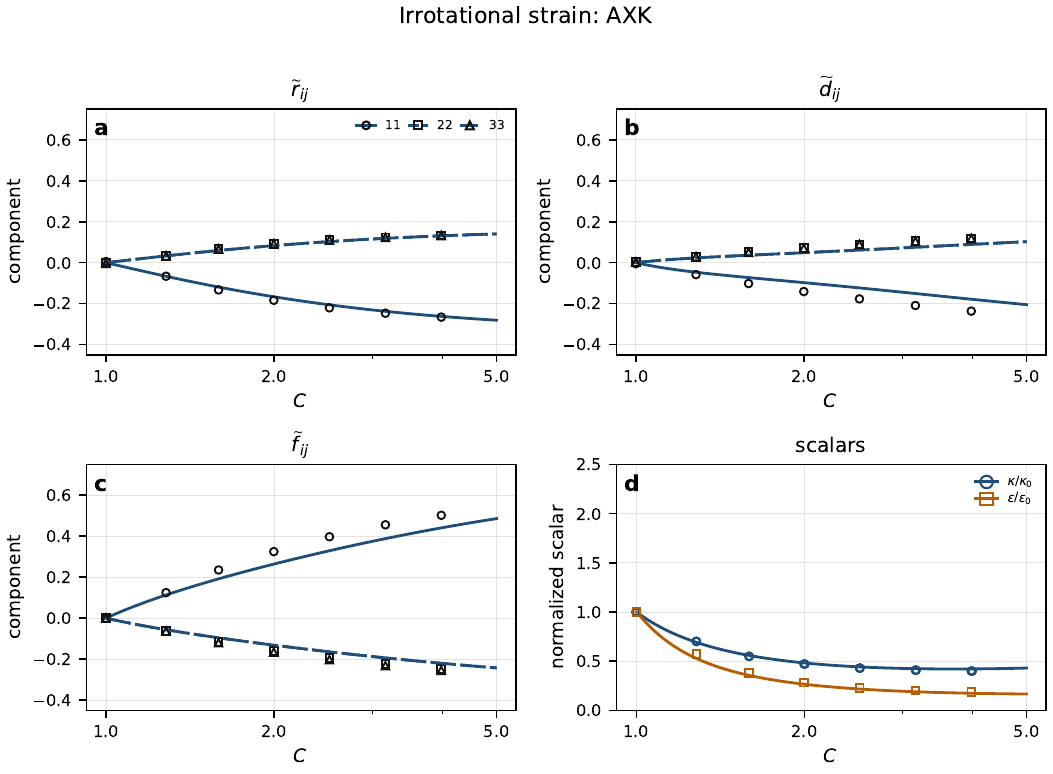}
  \caption{Atlas comparison for irrotational strain case AXK.  Tensor panels show the diagonal components of \(\widetilde r_{ij}\), \(\widetilde d_{ij}\), and \(\widetilde f_{ij}\); panel d shows \(\kappa/\kappa_0\) and \(\epsilon/\epsilon_0\) on the common scalar range used throughout the appendix. Lines denote predictions of the adopted RC--IPRM closure; open symbols show comparison data reported by Lee and Reynolds~\citep{LeeReynolds1985}.}
  \label{fig:atlas-AXK}
\end{figure}

\begin{figure}[!htbp]
  \centering
  \includegraphics[width=0.93\textwidth]{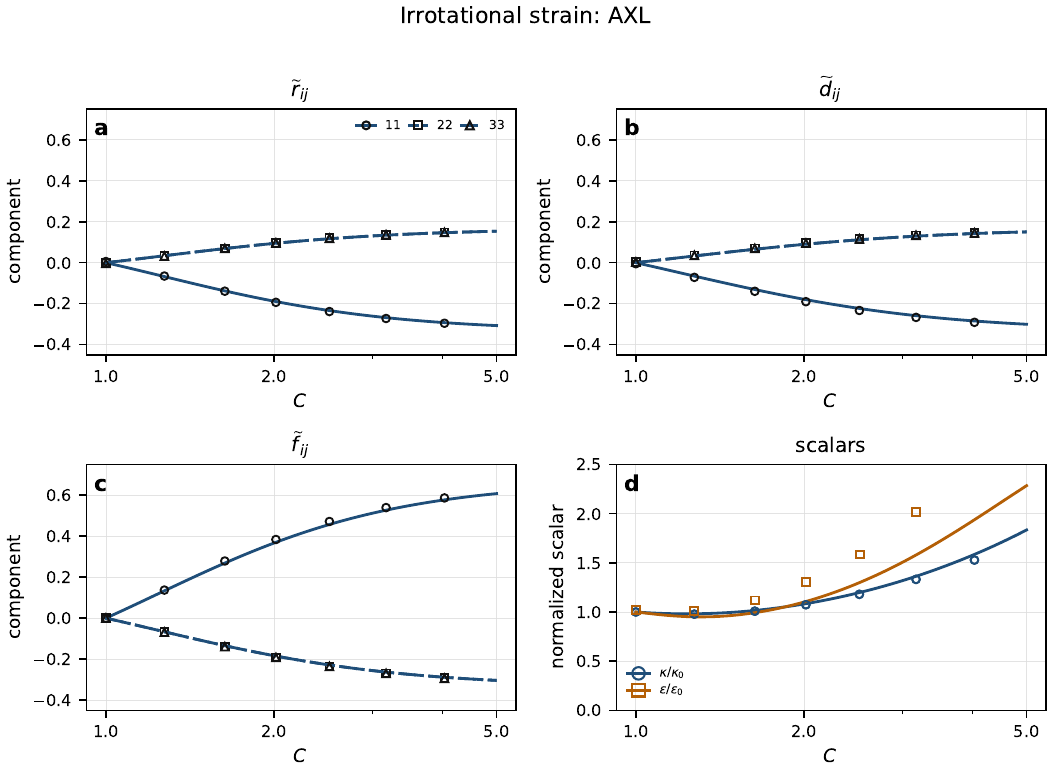}
  \caption{Atlas comparison for irrotational strain case AXL.  This case is one of the representative slow-strain examples discussed in Fig.~\ref{fig:lse-correction-strain}; the atlas view adds the corresponding componentality, dimensionality, and circulicity tensors. Lines denote predictions of the adopted RC--IPRM closure; open symbols show comparison data reported by Lee and Reynolds~\citep{LeeReynolds1985}.}
  \label{fig:atlas-AXL}
\end{figure}

\begin{figure}[!htbp]
  \centering
  \includegraphics[width=0.93\textwidth]{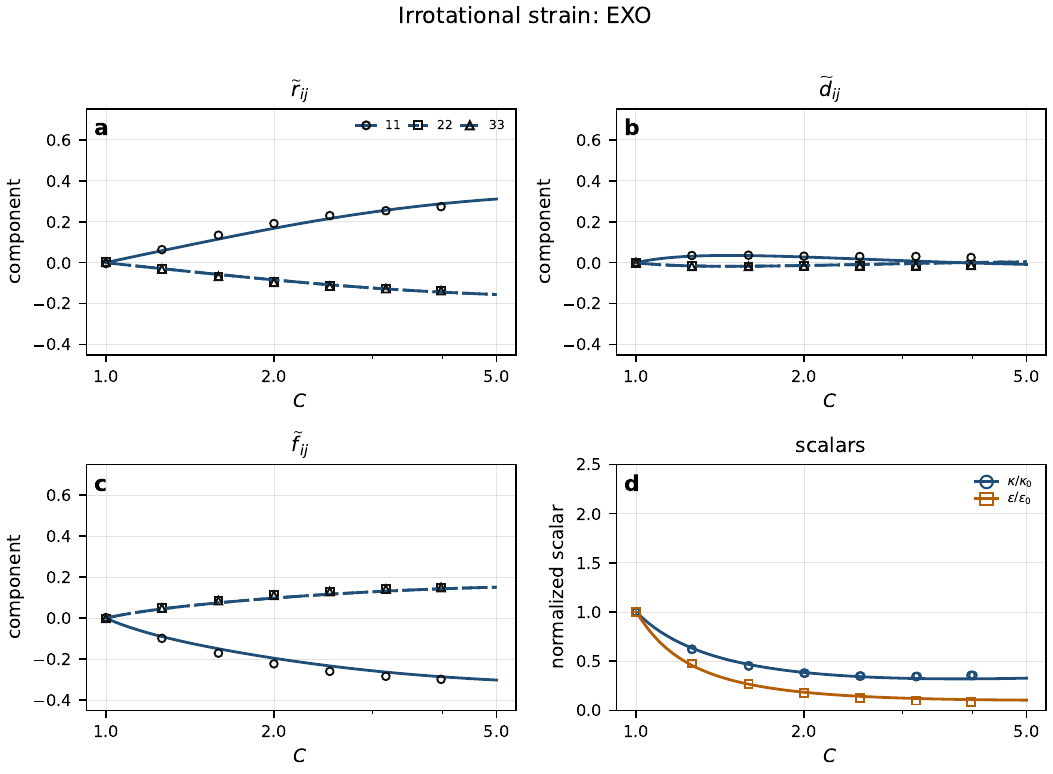}
  \caption{Atlas comparison for irrotational strain case EXO.  The same reference closure and plotting convention are used as in Fig.~\ref{fig:atlas-AXK}. Lines denote predictions of the adopted RC--IPRM closure; open symbols show comparison data reported by Lee and Reynolds~\citep{LeeReynolds1985}.}
  \label{fig:atlas-EXO}
\end{figure}

\begin{figure}[!htbp]
  \centering
  \includegraphics[width=0.93\textwidth]{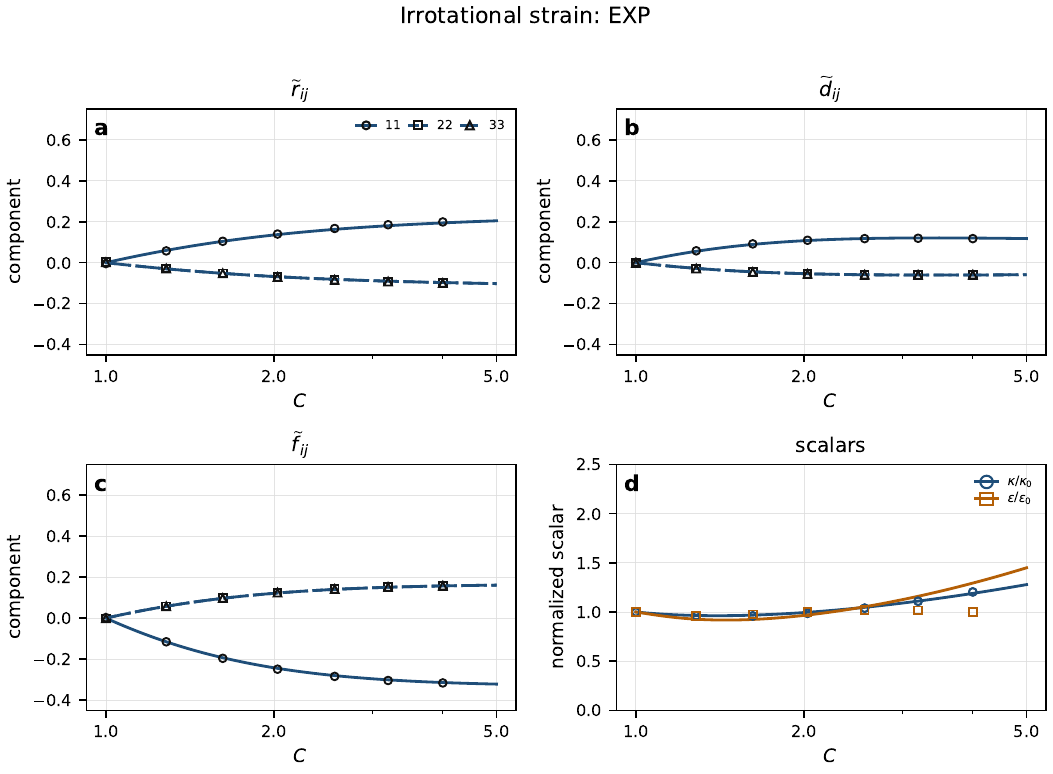}
  \caption{Atlas comparison for irrotational strain case EXP.  The common scalar axis in panel d is retained to make the energy and dissipation histories directly comparable with the other strain modes. Lines denote predictions of the adopted RC--IPRM closure; open symbols show comparison data reported by Lee and Reynolds~\citep{LeeReynolds1985}.}
  \label{fig:atlas-EXP}
\end{figure}

\begin{figure}[!htbp]
  \centering
  \includegraphics[width=0.93\textwidth]{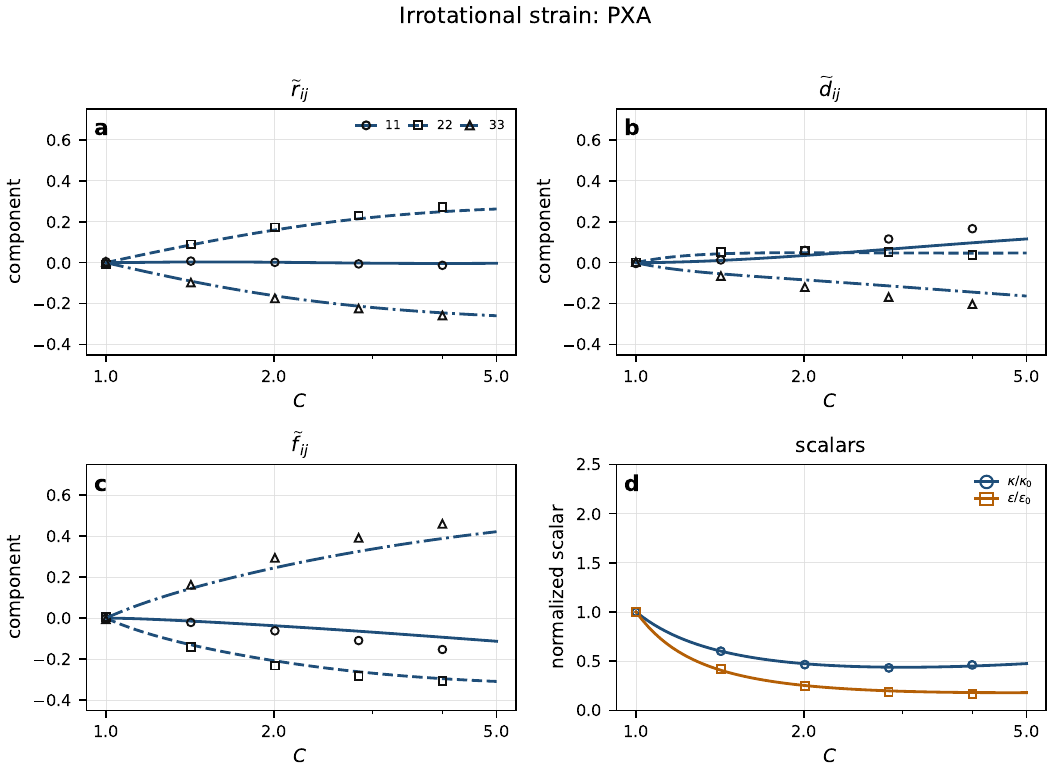}
  \caption{Atlas comparison for irrotational strain case PXA.  The tensor panels provide the structure-tensor comparison corresponding to the same reference closure and projection density used throughout the strain atlas. Lines denote predictions of the adopted RC--IPRM closure; open symbols show comparison data reported by Lee and Reynolds~\citep{LeeReynolds1985}.}
  \label{fig:atlas-PXA}
\end{figure}

\begin{figure}[!htbp]
  \centering
  \includegraphics[width=0.93\textwidth]{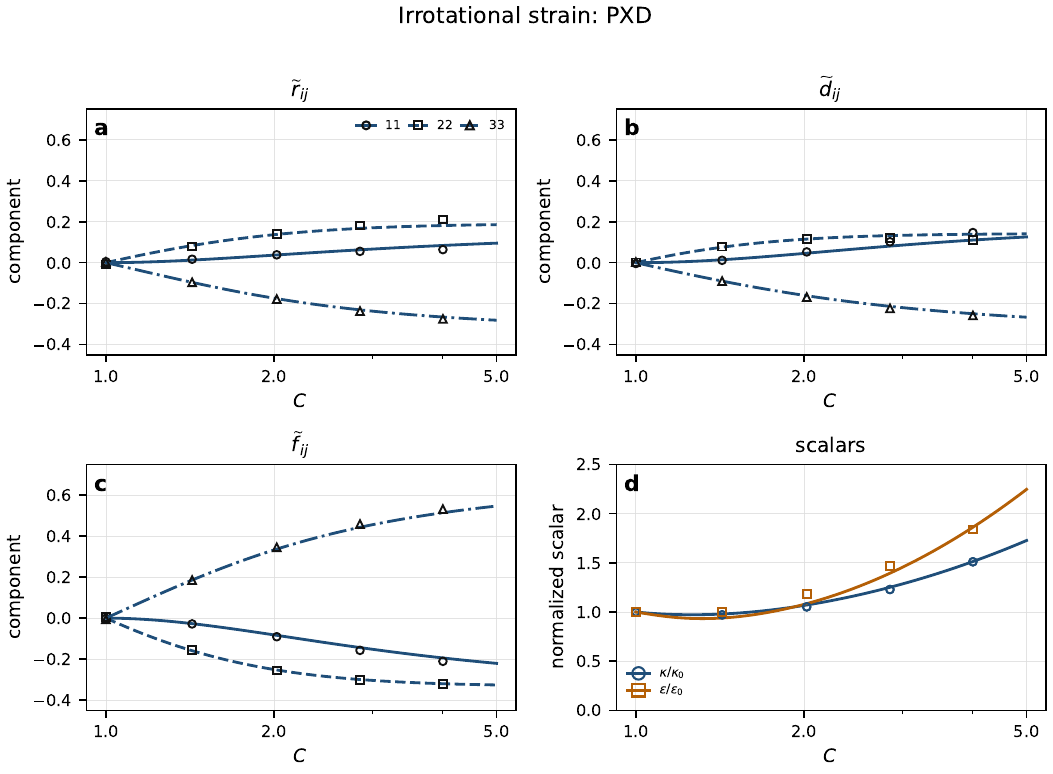}
  \caption{Atlas comparison for irrotational strain case PXD.  This case is the second representative slow-strain example discussed in Fig.~\ref{fig:lse-correction-strain}; the atlas view records the full tensor comparison in addition to the scalar histories. Lines denote predictions of the adopted RC--IPRM closure; open symbols show comparison data reported by Lee and Reynolds~\citep{LeeReynolds1985}.}
  \label{fig:atlas-PXD}
\end{figure}

Figures~\ref{fig:atlas-AXK}--\ref{fig:atlas-PXD} show the tensor and
scalar Atlas comparisons with the irrotational-strain comparison data
reported by Lee and Reynolds~\citep{LeeReynolds1985} for the six cases
AXK, AXL, EXO, EXP, PXA, and PXD.  These figures extend the main-body
strain comparison by showing the component-level behavior of the
normalized componentality, dimensionality, and circulicity tensors,
together with the scalar energy and terminal-drain histories.

The figure convention follows the main text.  RC--IPRM predictions are shown by continuous lines, and comparison data are shown by open markers.  The tensor panels show the diagonal components 11, 22, and 33 of the deviatoric normalized componentality, dimensionality, and circulicity tensors, \(\widetilde r_{ij}\), \(\widetilde d_{ij}\), and \(\widetilde f_{ij}\).  The 12 components vanish identically for these irrotational-strain cases in the chosen coordinate system and are omitted to keep the atlas figures uncluttered.  The scalar panel uses a common vertical range, \(0\le \kappa/\kappa_0,\epsilon/\epsilon_0\le 2.5\), in all six cases so that differences among strain modes can be judged directly.  The Lee--Reynolds tensor data are reported in deviatoric normalized form and are plotted consistently on the \(\widetilde r\), \(\widetilde d\), and \(\widetilde f\) panels.

\end{document}